\documentclass[preprint,preprintnumbers,amsmath,amssymb,12pt,floatfix,epsfig,nofootinbib,superscriptaddress]{revtex4}

\usepackage{amsmath} 
\usepackage{bm}      
\usepackage[usenames, dvipsnames]{color}
\usepackage{graphicx}
\usepackage{amssymb}
\usepackage[normalem]{ulem}
\renewcommand\sout{\bgroup\color[rgb]{1,0,0} \ULdepth=-.5ex \ULset}

\begin{document}
\title{Nuclear shape evolution of neutron-deficient Au and kink structure of Pb isotopes}
\author{Myeong-Hwan Mun} 
\affiliation{Department of Physics and Origin of Matter and Evolution of Galaxies (OMEG) Institute, Soongsil University, Seoul 06978, Korea}

\author{Eunja Ha \footnote{ejaha@hanyang.ac.kr}}
\address{Department of Physics and Research Institute for Natural Science, Hanyang University, Seoul, 04763, Korea}

\author{Yong-Beom \surname{Choi}}
\affiliation{Center for Innovative Physicist Education and Research, Pusan National University, Busan 46241, Korea}

\author{Myung-Ki Cheoun}
\email{cheoun@ssu.ac.kr (Corresponding Author)}
\affiliation{Department of Physics and Origin of Matter and Evolution of Galaxies (OMEG) Institute, Soongsil University, Seoul 06978, Korea}

\date{\today}

\begin{abstract}
Recent experiments using advanced laser spectroscopy technique revealed that the charge radii of neutron-deficient gold (Au) isotopes exhibit significant changes in ground state deformation: odd-even shape staggering in the $N = 98 \sim 100$ region and abrupt change of charge radii from $N =$ 108. In this study, we examine the abnormal shape evolution of the nuclear charge radii. To understand the nuclear structure underlying this phenomenon, we exploit the deformed relativistic Hartree–Bogoliubov theory in continuum (DRHBc). The significant change in mean-squared charge radii ($\delta {< r^2 >}$) turns out to originate from nuclear shape transitions between prolate deformation and small oblate deformation due to the shape coexistence possibility. We elucidate the nuclear shape evolution by analyzing the evolution of occupation probability for single-particle states. In addition, the abrupt kink structure in the nuclear charge radius of lead (Pb) isotopes near the $N =$ 126 shell is also investigated and reproduced quite well.
\end{abstract}

\maketitle

\section{Introduction}

The nuclear charge radius is one of the most important properties that can be direct evidence of  the fundamental characteristics of nuclear structure. Furthermore, it is significant in investigating the evolution of nuclear structures, including the emergence of new magic numbers or the disappearance of traditional magic numbers \cite{Angeli2013, Angeli2015, Wang2013, Gorges2019}. In particular, the nuclear charge radius could provide critical information for shape transition and coexistence \cite{Wood1992, Cejnar2010,Witte2007,Flanagan2013,Bao2021} as well as isomerism \cite{Geithner2008,Nort2009,Yordanov2016}.

If a constant saturation density is assumed inside the nucleus, the nuclear charge radius is typically described by the $A^{1/3}$ law \cite{Bohr,Duflo1994,Die2009,Qian2014,Ma2017,Li2021} (where $A$ represents the mass number). In recent years, with the development of experimental technology, the elative root-mean-square (rms) charge radii of unstable nuclei have been increasingly measured by various approaches such as ion guide isotope separateor \cite{Vries1987,Avgo2011}, $K_{\alpha}$~X-ray isotope shifts \cite{Mari2015,Angeli2016,Boehm1974,Mano2019,Han2022}, and high-resolution laser spectroscopy \cite{Krie2012,Seli2013,Garc2016,Mina2016,Marsh2018,Miller2019,Kauf2020,Groote2020,Cubiss2023}. 
These measurements of the hyperfine interaction have been performed on a variety of isotopes, ranging from K and Ca to Au, Pb, and Bi isotopes \cite{Rose2015,Krei2014,Kosz2019,Kosz2021,Tanaka2020,Wien2013,Geld2022,Gorg2019,Good2021,Mars2018,Kuehl1977,Ulm1986,Anse1986,Barz2021,Cubiss2023}.
Experimental investigations of nuclear charge radii have found meaningful regular patterns, abrupt changes, and non-linear trends along isotopic chains in the nuclear chart. Most recently, the strong odd–even shape staggering (OES) in some isotopic chains and the abrupt kinks across neutron shell closures have been observed and attracted interest \cite{Garc2016,Miller2019,Groote2020,Good2021,Kosz2021,Geld2022,Godd2013,Hamm2018,Repo2021,Malb2022}.

A more accurate description of nuclear charge radii, which are closely associated with nuclear deformation, is necessary for understanding these nuclear shape transition. In theory, a variety of methods, such as local-relationship-based models \cite{Sun2017,CMa2021}, macroscopic-microscopic models \cite{Buch2005,Iimu2008}, non-relativistic mean-field models \cite{Gori2010,Rein2017}, relativistic mean-field models \cite{ADND2022,Xia2018,Pere2021}, and ab initio no-core shell models \cite{Fors2009,Chou2020} are employed for systematic investigation of nuclear charge radii. Recently, on account of the development of high-performance computing, several machine learning approaches are also widely used to improve the description of nuclear charge radii \cite{YFMa2020,Wu2020,JQMa2022,Dong2023}.
Although each model provides fairly good descriptions of the nuclear charge radii across the nuclear chart, only a few of these models can accurately reproduce strong odd-even shape staggering (OES) and abrupt kinks across the neutron shell closure \cite{Mun2023,Sels2019}.

In this report, we demonstrate that the observed shape transition could be attributed to the shape coexistence of the Au and Pb isotopes using the deformed relativistic Hartree–Bogoliubov theory in continuum (DRHBc) \cite{Zhou2010,Li2012,Lulu2012}. Firstly, we present the shape coexistence of Au (Pb) isotopes with a precision of 1 MeV (1.5 MeV) using total binding energy curves. Secondly, we argue that the odd-even shape staggering in the $N = 98 \sim 100$ region and abrupt change $N =$ 108 are closely associated with the shape coexistence. Finally, we affirm that the kink structure in the vicinity of $^{208}$Pb is linked to the increase of neutron radii for Pb isotopes.
This article is arranged as follows. A brief description of the DRHBc theory used in the present calculation is presented in Sec. II. Detailed results of Au and Pb isotopes including the nuclear shape transition and kink are provided in Sec. III. Finally, the summary and conclusion are given  in Sec. IV.

\section{Formalism}

In order to comprehensively address the aforementioned issues, well-refined and state-of-the-art relativistic nuclear model is indispensable required. This model should simultaneously consider deformation, pairing correlations, and the continuum effects within a microscopic framework capable of encompassing the entire nuclear mass range. Regarding this issue, the deformed relativistic Hartree–Bogoliubov theory in continuum (DRHBc), based on meson-exchange \cite{Zhou2010,Lulu2012,Li2012} or point-coupling \cite{Kai2020,Cong2022} density functionals, has been developed. More lately, the DRHBc theory has been successfully extended to deal with odd-mass or odd-odd nuclei by employing an automatic blocking method \cite{Li2012,Cong2022}. 
The DRHBc theory has been proven to offer a strong description of nuclear masses with high predictive power \cite{Kaiyuan2021,Pan2021,ADND2022,Peng2024}, and it has demonstrated its capabilities through various applications \cite{Cong2019,Sun2018,Sun2020,In2021,Yang2021,Sun2021,Sun2021-2,Sun2021-3,Kim2022, Zhang2019, Zhang2023, ZhangPLB, An2024}. For example, the DRHBc theory has been successfully applied to shape staggering and kink structure in the Hg isotopes \cite{Mun2023}, as well as the shape coexistence and prolate dominance in the Te, Xe, and Ba isotopes \cite{Peng2023}.

In this study, we concentrate on shape coexistence and its subsequent effects, specifically nuclear shape transition, i.e., odd-even shape staggering and abrupt shape change in Au isotopes, and kink structure in Pb isotopes employing the DRHBc theory. The details of the DRHBc theory can be found in Refs. \cite{Zhou2010,Lulu2012,Li2012,Kai2020,Cong2022}. Here, only a brief introduction to the formalism of the DRHBc theory is provided. In the DRHBc theory, the relativistic Hartree-Bogoliubov (RHB) equation  \cite{Kucharek1991}, which is solved in a Dirac Woods-Saxon basis \cite{Zhou2003,Dirac2022}, is given as follows

\begin{equation} \label{eq:hfbeq}
\left( \begin{array}{cc} h_D - \lambda &
\Delta  \\
 - \Delta^{*} & - h^{*}_D + \lambda
  \end{array}\right)
\left( \begin{array}{c}
U_{k} \\ V_{k}  \end{array}\right)
 =
 E_{k}
\left( \begin{array}{c} U_{k} \\
V_{k} \end{array}\right),
\end{equation}
where $\lambda$ is the Fermi energy and $E_k$ and $(U_k,V_k)^{T}$ are the quasiparticle energy and quasiparticle wave function. In coordinate space, the Dirac Hamiltonian $h_D$ is given by
\begin{equation} \label{eq:hd}
h_D = {\bm {\alpha}} \cdot {\bm p}  +  V(\bm{r})  +  \beta [M + S({\bm r})] , 
\end{equation}
where $M$ is the nucleon mass, and $V(\bm{r})$ and $S(\bm{r})$ are the vector and scalar potentials, respectively.

The paring potential $\Delta$ is given with the pairing tensor $\kappa ({\bf r}, {\bf r}^{'})$ \cite{Ring2004} as follows
\begin{equation}
\Delta({\bf r}, {\bf r}^{'}) = V ({\bf r}, {\bf r}^{'}) \kappa ({\bf r}, {\bf r}^{'})
\end{equation}
with a density-dependent zero range force
\begin{equation}
V({\bf r}, {\bf r}^{'}) = {V_0 \over 2} ( 1 - P_{\sigma}) \delta ( {\bf r} - {\bf r}^{'}) ( 1 -  { \rho (\bf r) \over \rho_{sat}} )~.
\end{equation}
where $\rho_{sat}$ is the nuclear saturation density, $V_0$ is the pairing strength, and $(1-P_{\sigma})$ is the projector for the spin $S = 0$ component in the pairing channel.

For an axially deformed nucleus with spatial reflection symmetry, the potentials and densities can be expanded in terms of Legendre polynomials \cite{Price1987},

\begin{equation} \label{eq:lambda}
f(\bm{r}) = \sum_{\lambda} \, f_{\lambda} (r) P_{\lambda} (\rm{cos}\theta),
\,\,\, \lambda = 0, \, 2, \, 4, \, \cdots.
\end{equation}

To determine the ground state of an odd-mass or odd-odd isotopes, consideration should be given to the blocking effect of the unpaired nucleon(s) \cite{Ring2004}. For this purpose, the DRHBc theory appropriately handles the blocking effect of unpaired nucleon(s) by employing either the orbital-fixed or automatic blocking procedure in odd nuclei. In the current DRHBc theory, the equal filling approximation \cite{Li2012, Martin2008}, which preserves time-reversal symmetry and circumvents computationally intensive calculations, is commonly adopted. In Ref. \cite{Cong2022}, it is confirmed that the DRHBc theory, which incorporates the blocking effect based on the point-coupling functional, is capable of accurately describing odd-mass and odd-odd nuclei. Remarkably, the ground states obtained by the automatic blocking for nuclei such as $^{22}$Al and  $^{23}$Mg were found to be identical to those using the orbital-fixed blocking \cite{Cong2022}. This indicates the reliability and consistency of the automatic blocking method in the current numerical calculations.

For the numerical calculations of Au and Pb isotopes, we employ a pairing strength of $V_0$ = -- 325.0 MeV fm${^3}$, with a pairing window of 100 MeV, and adopt a saturation density of $\rho_{sat}$ = 0.152 fm$^{-3}$. The energy cutoff $E_{cut}^+ =$ 300 MeV and the angular momentum cutoff $J_{\max} = (23/2) \hbar $ are taken for the Dirac Woods-Saxon basis. In Eq. (\ref{eq:lambda}), the Legendre expansion truncation is chosen as $\lambda_{\max}=8$. The above numerical details are the same as those used in the global DRHBc mass table calculations over the nuclear chart \cite{Kai2020,Kaiyuan2021,Cong2022}.


\section{Results}

\begin{figure}
\centering
\includegraphics[width=0.46\linewidth]{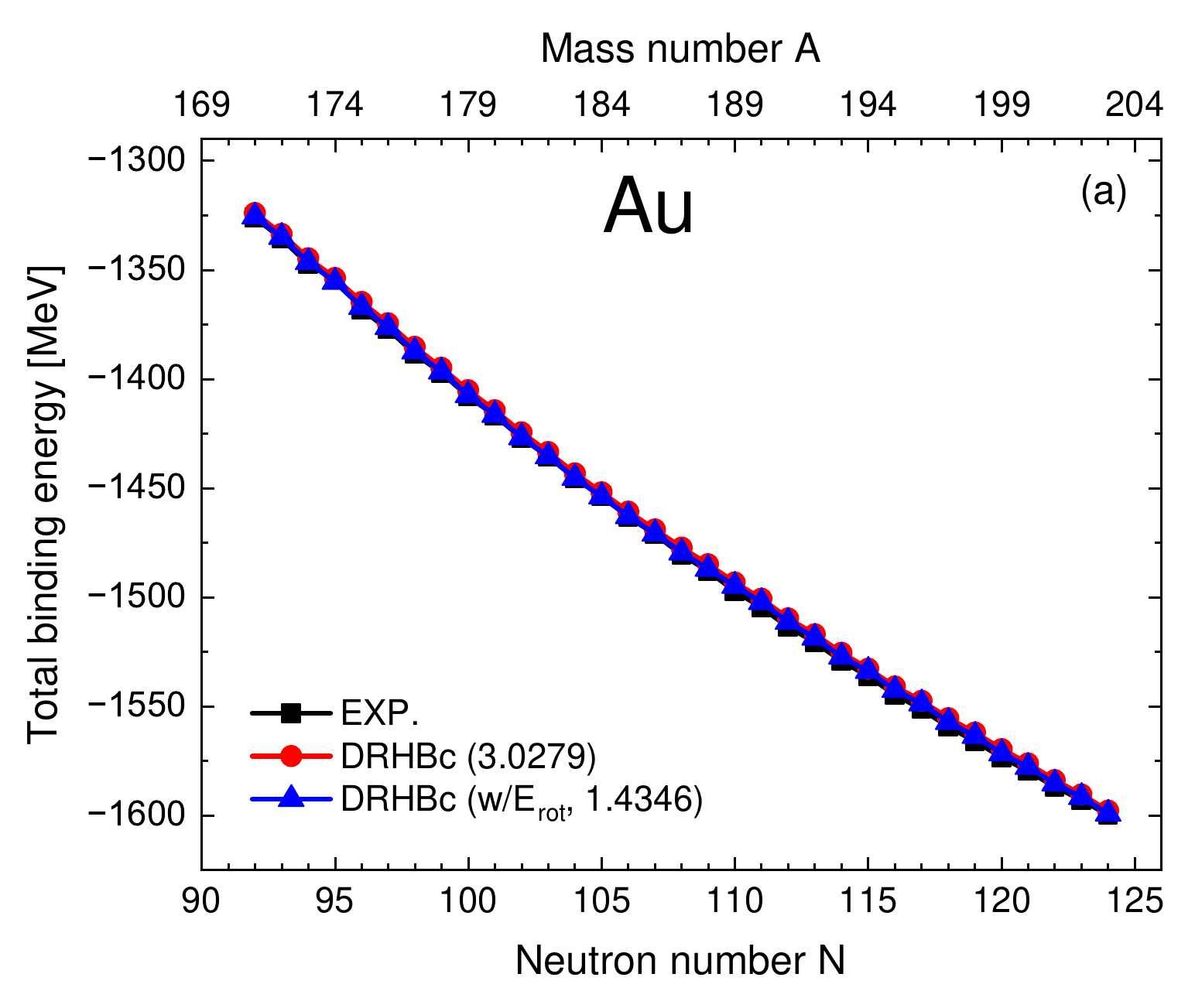}
\includegraphics[width=0.45\linewidth]{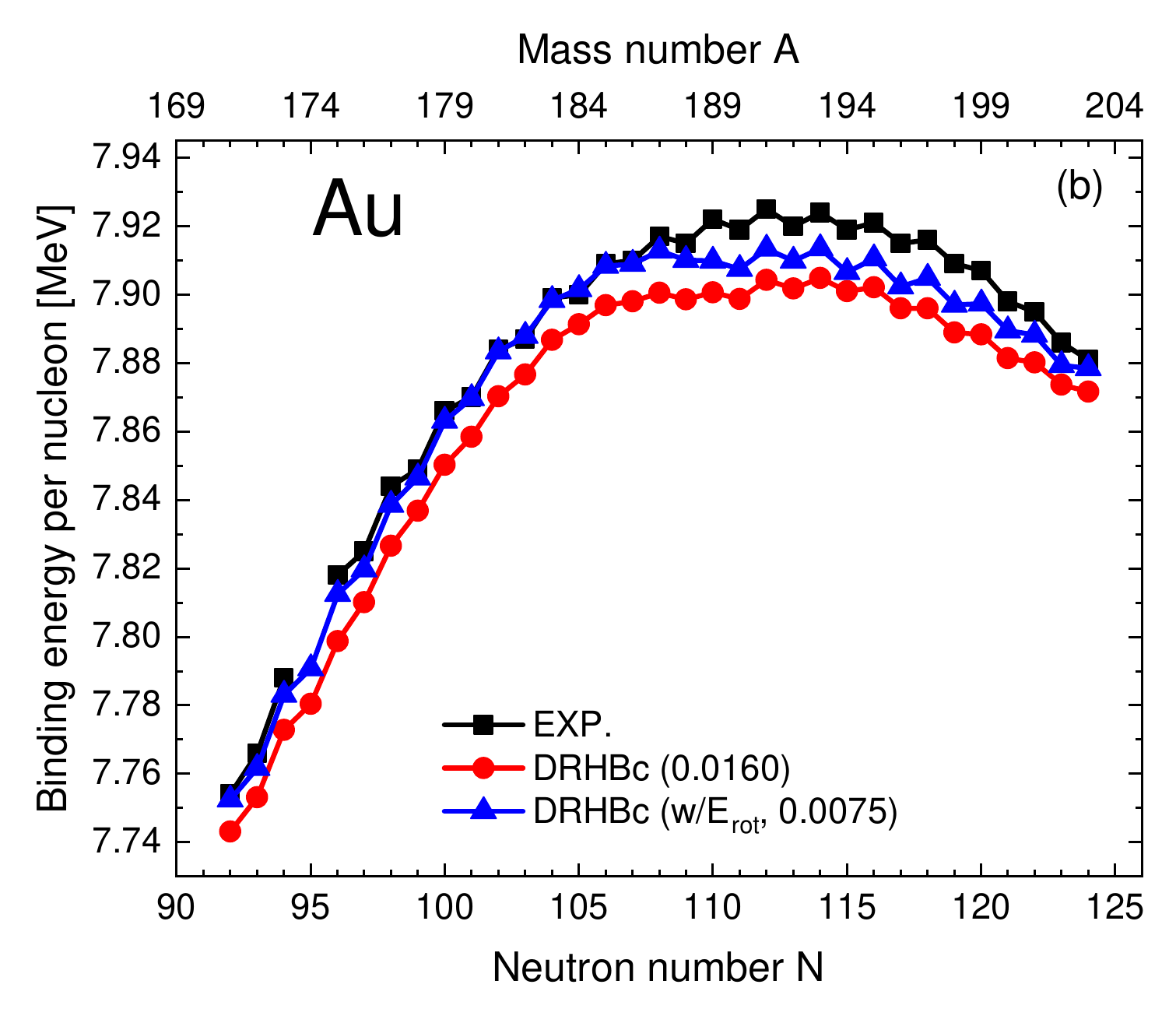}
\includegraphics[width=0.46\linewidth]{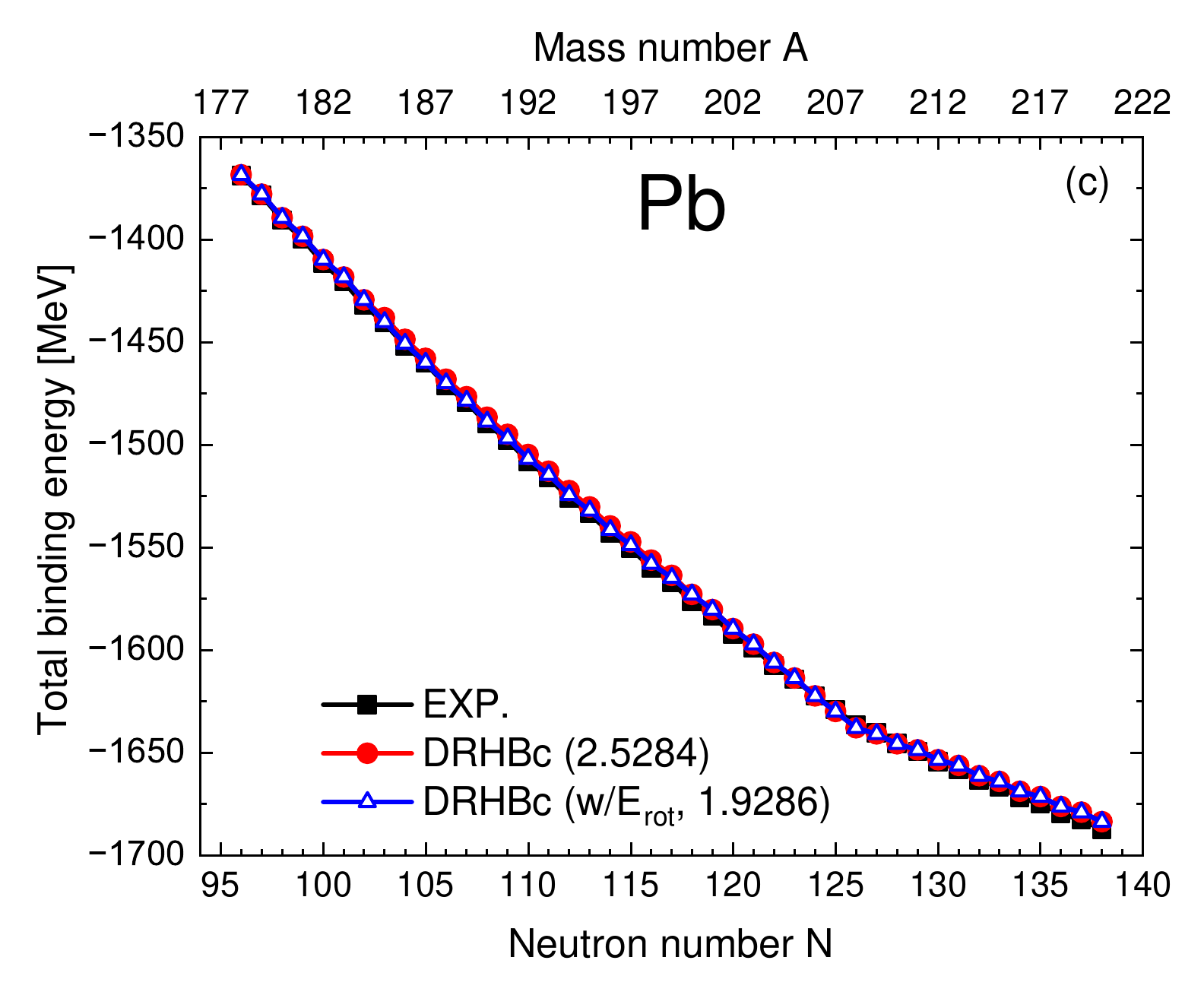}
\includegraphics[width=0.46\linewidth]{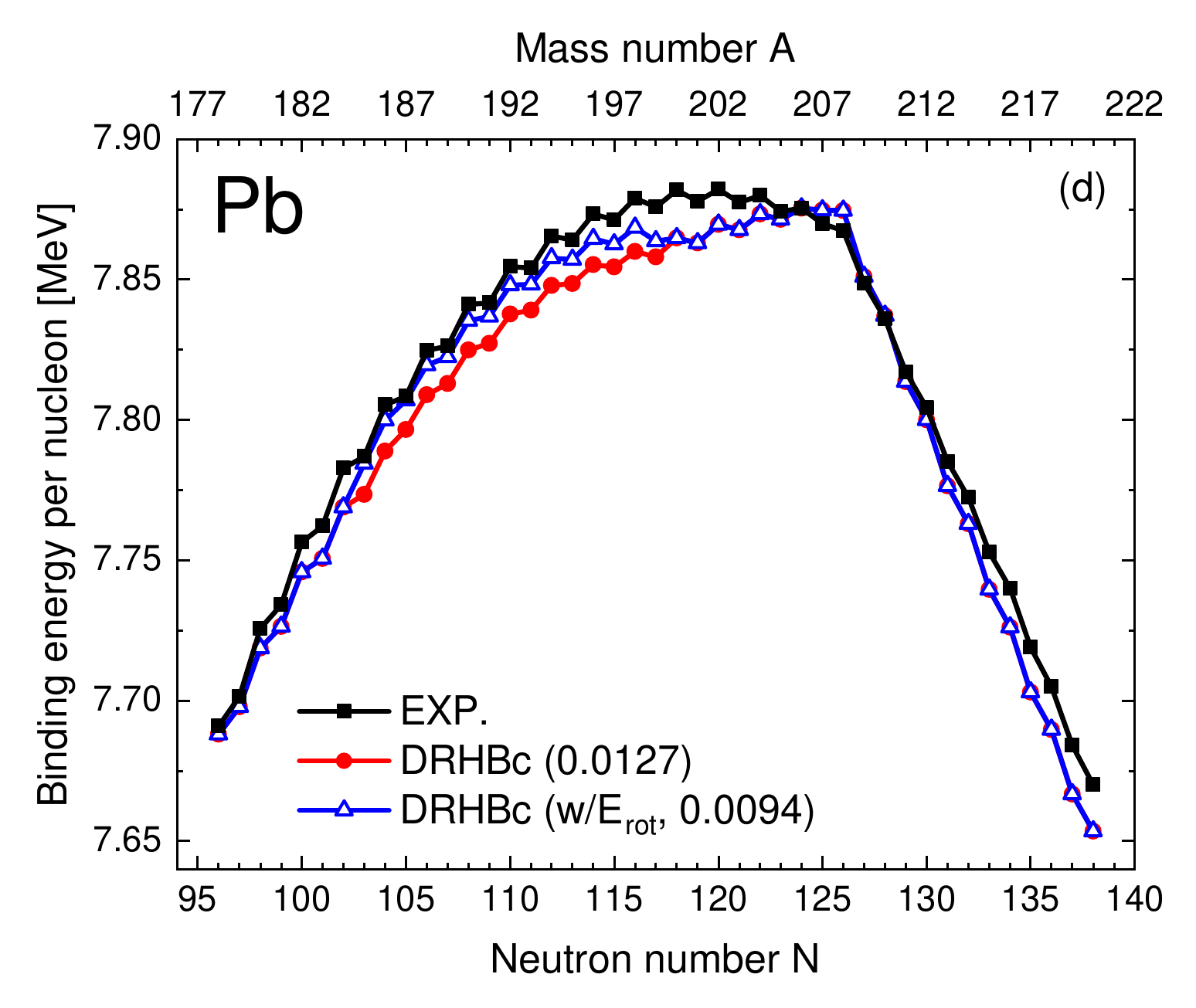}
\caption{(Color online) TBEs (a) ((c)) and BE/A (b) ((d)) for Au (Pb) isotopes determined by DRHBc calculations without (red circles) and with (blue triangles) rotational correction $E_\mathrm{rot}$. They are compared with available experimental data taken from Ref. \cite{AME2020}. The numbers in parentheses stand for average root-mean-square deviation in MeV to the data \cite{AME2020}.}
\label{fig1}
\end{figure}

\begin{figure}
\centering
\includegraphics[width=0.49\linewidth]{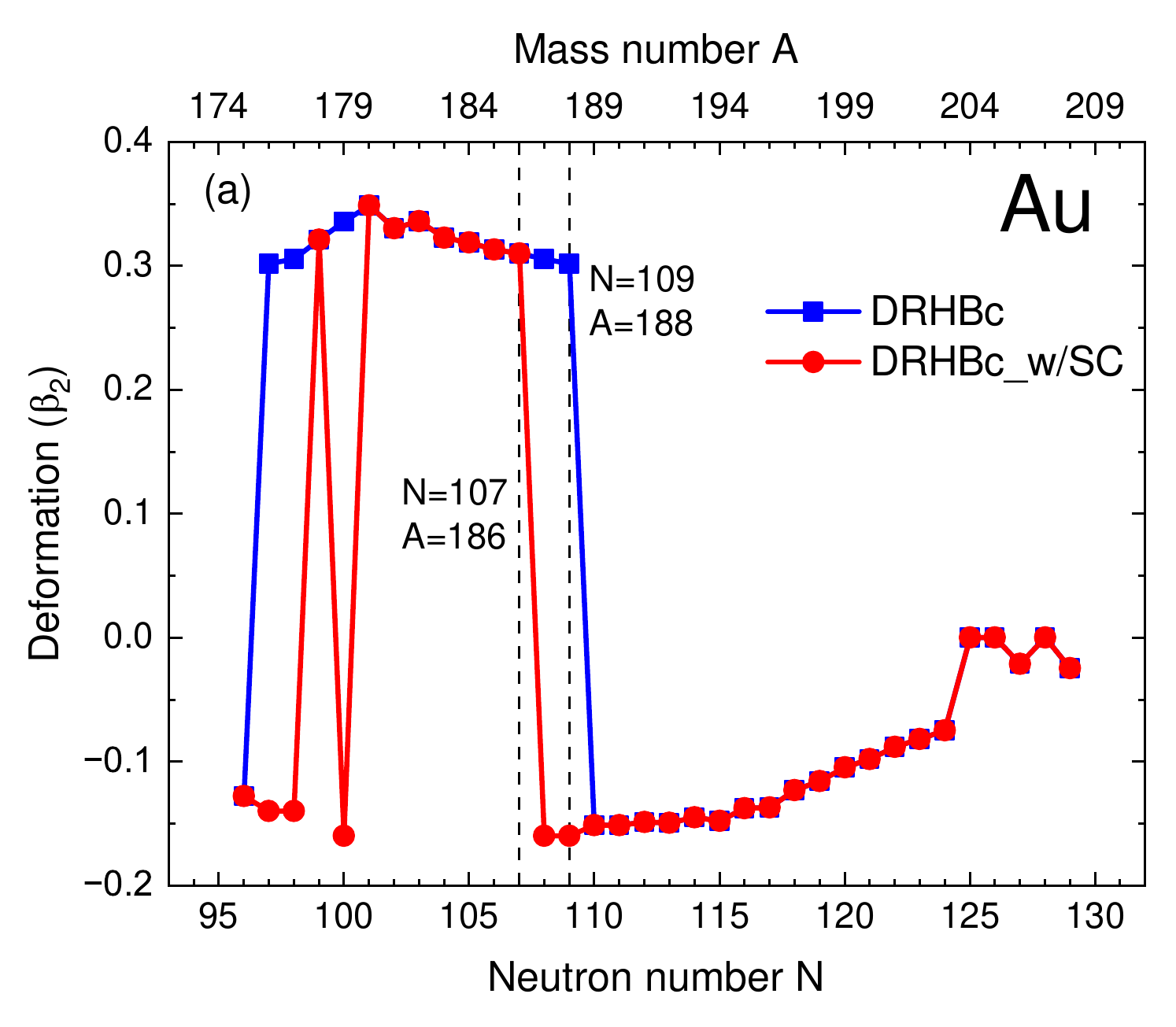}
\includegraphics[width=0.49\linewidth]{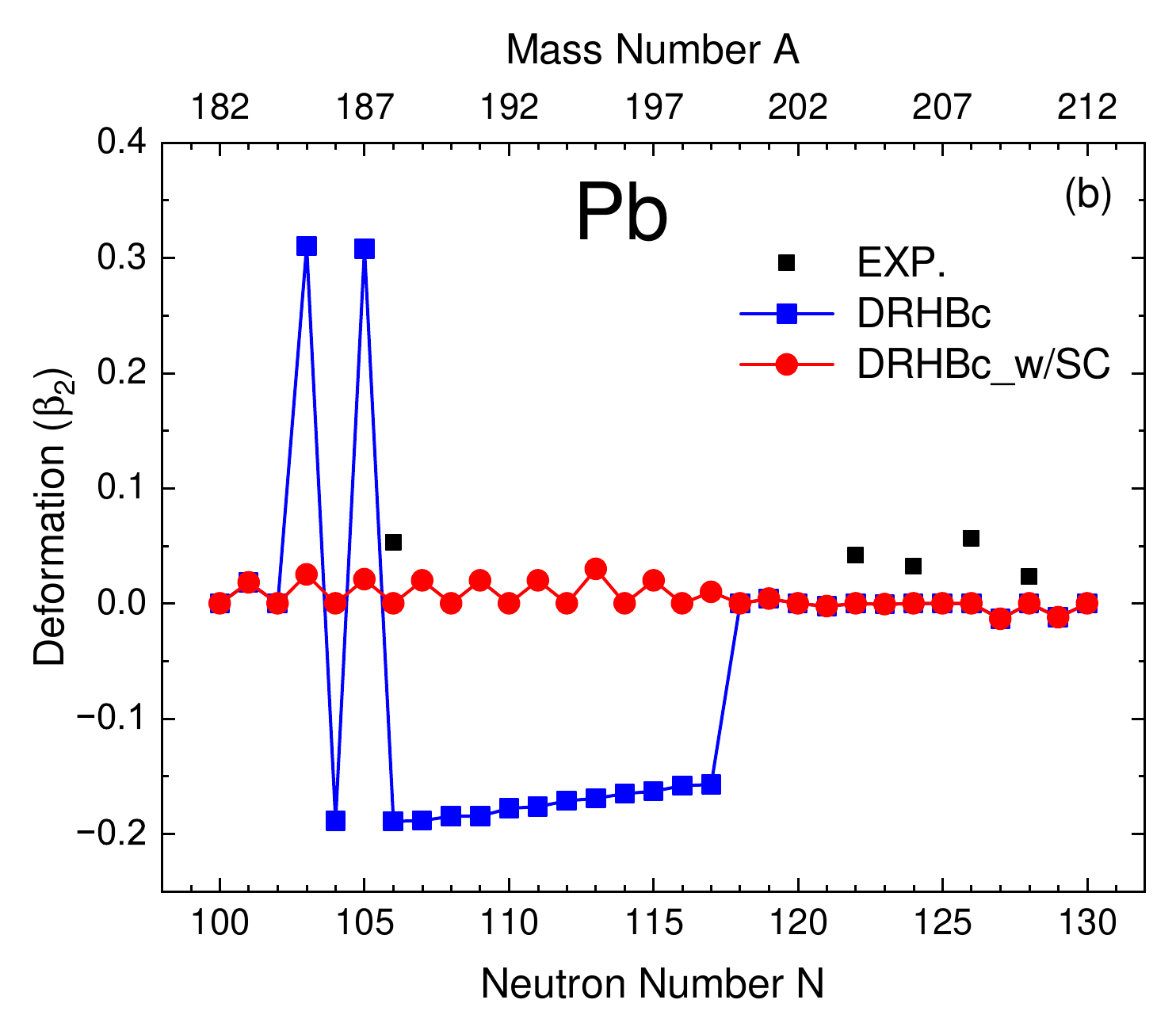}
\caption{(Color online) Quadrupole deformation parameters $\beta_2$ of Au (a) and Pb (b) isotopes by DRHBc mass model with available experimental data from Ref. \cite{NNDC}. Results by blue color are obtained by DRHBc calculations, while those by red color are obtained by considering the shape coexistence (see the text for details).}
\label{fig2}
\end{figure}
Firstly, we examined the total binding energy (TBE) and binding energy per nucleon (BE/A) of Au and Pb isotopes by the DRHBc theory. In Fig. \ref{fig1} , the TBEs (a) and BE/A (b) for Au and TBEs (c) and BE/A (d) for Pb obtained from the DRHBc calculations are shown versus the neutron number together with available experimental data \cite{AME2020}. Both the TBEs and BE/A demonstrate excellent agreement with the experimental data, showing uncertainties by root-mean-square deviation of less than 1.43 MeV and 0.01 MeV (1.93 MeV and 0.01 MeV) in the TBE and BE/A of Au (Pb), respectively. This result serves as a physical guideline for investigating the nuclear shape evolution in the Au isotopes and kink structure in the Pb isotopes.

The ground-state quadrupole deformations $\beta_2$ calculated by DRHBc are presented in Fig. \ref{fig2}. For gold isotopes (Fig. \ref{fig2} (a)), the prolate deformation is preferred to the oblate deformation in the region $A \leq 188$ (see the results of blue color). The transition from prolate deformations ($\beta_2 < 0.3$) to oblate deformations ($-0.1 < \beta_2 < -0.2$) occurs abruptly at $N = 109$ ($^{188}$Au). As shown in the blue color results of Fig. \ref{fig2} (b), the majority of lead isotopes within the $185 \leq A \leq 199$ range (excluding $^{185, 187}$Pb) favor the oblate deformations ($\beta_2 \simeq -0.2$). Additionally, there is an abrupt transition from oblate deformations ($\beta_2 < -0.2$) to nearly spherical shapes ($\beta_2 \simeq 0$) observed at $N = 118$ ($^{200}$Pb). The red color results are calculated by considering the theoretical shape coexistence, as will be explained later on.

\begin{figure}
\centering
\includegraphics[width=0.49\linewidth]{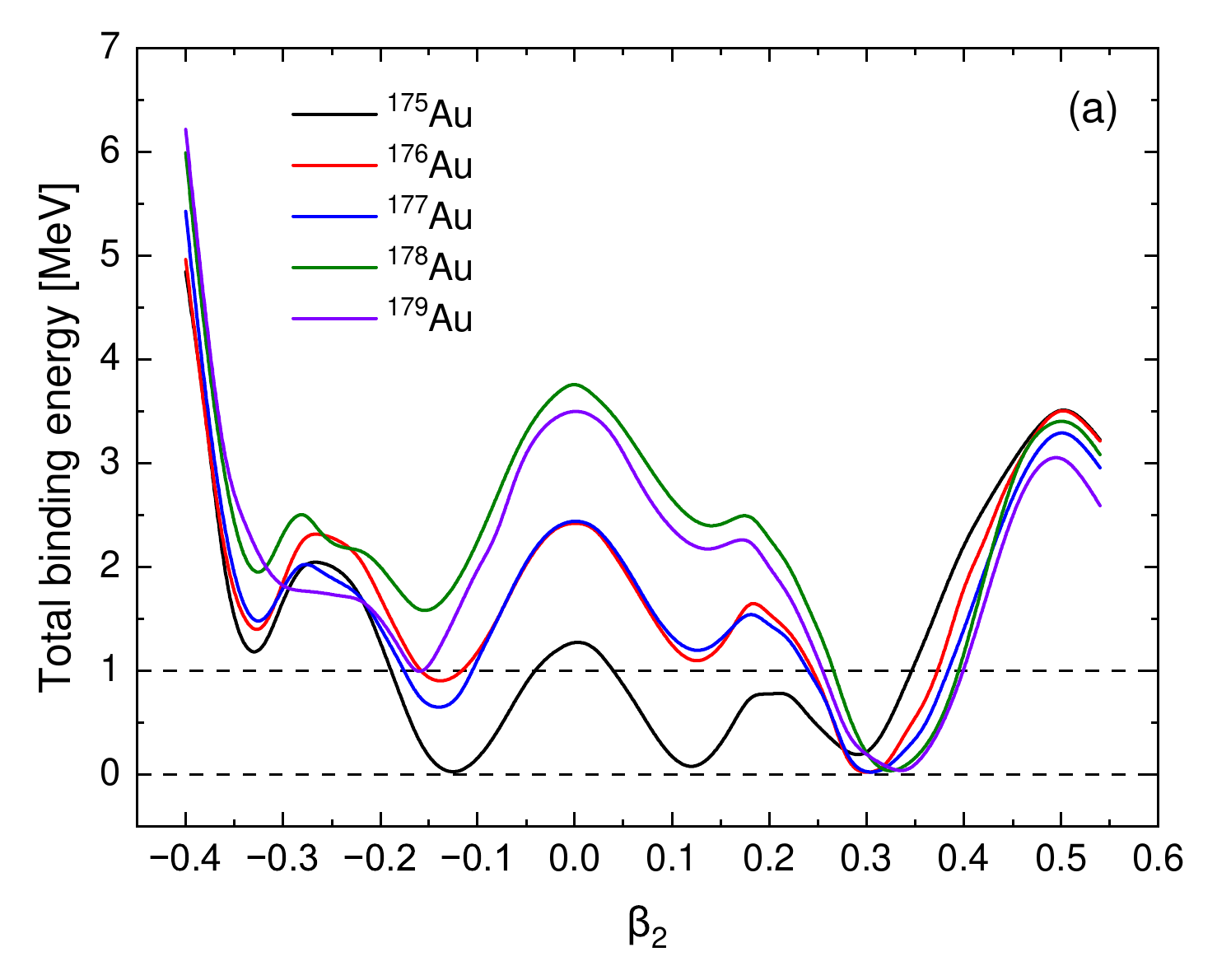}
\includegraphics[width=0.49\linewidth]{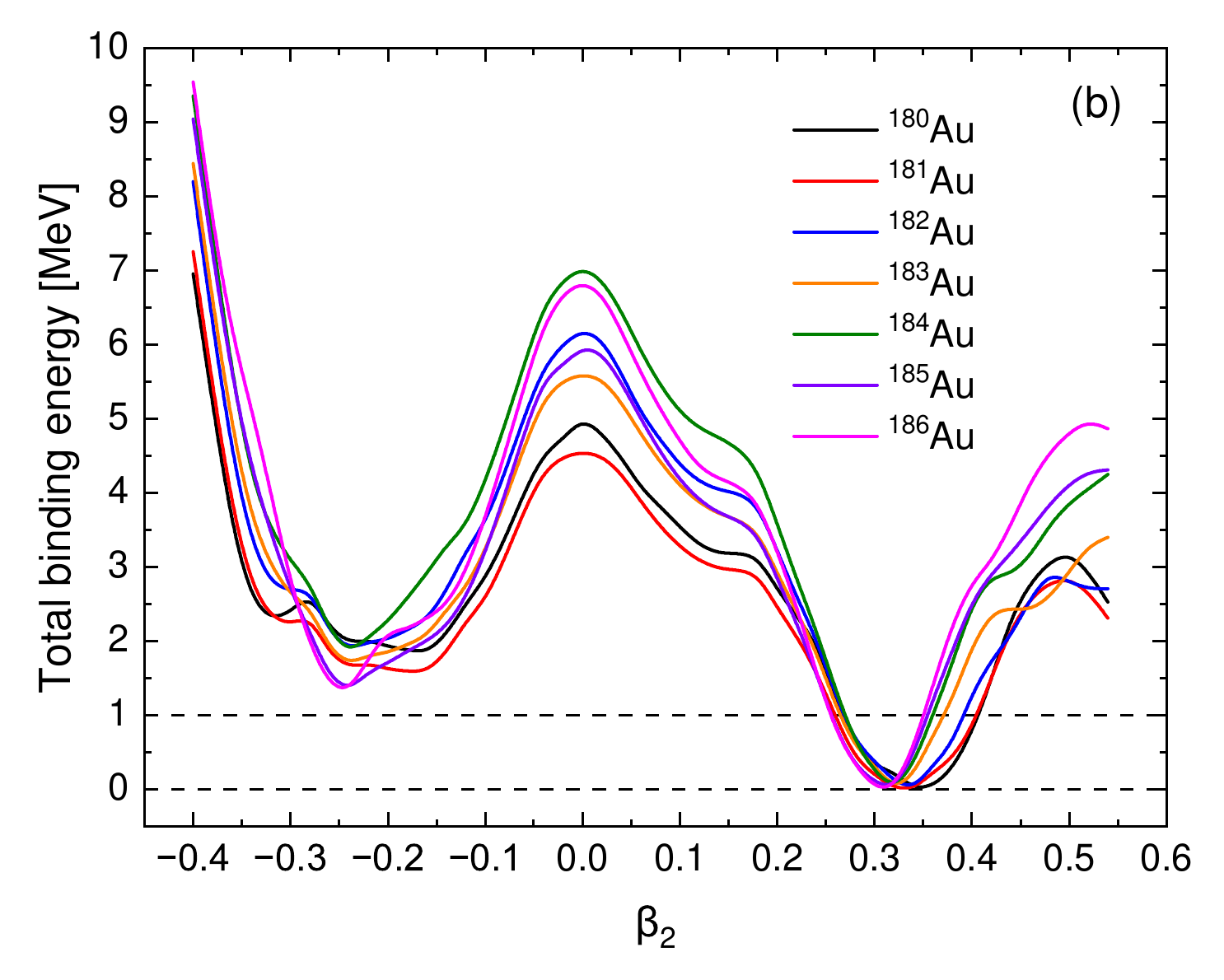}
\includegraphics[width=0.49\linewidth]{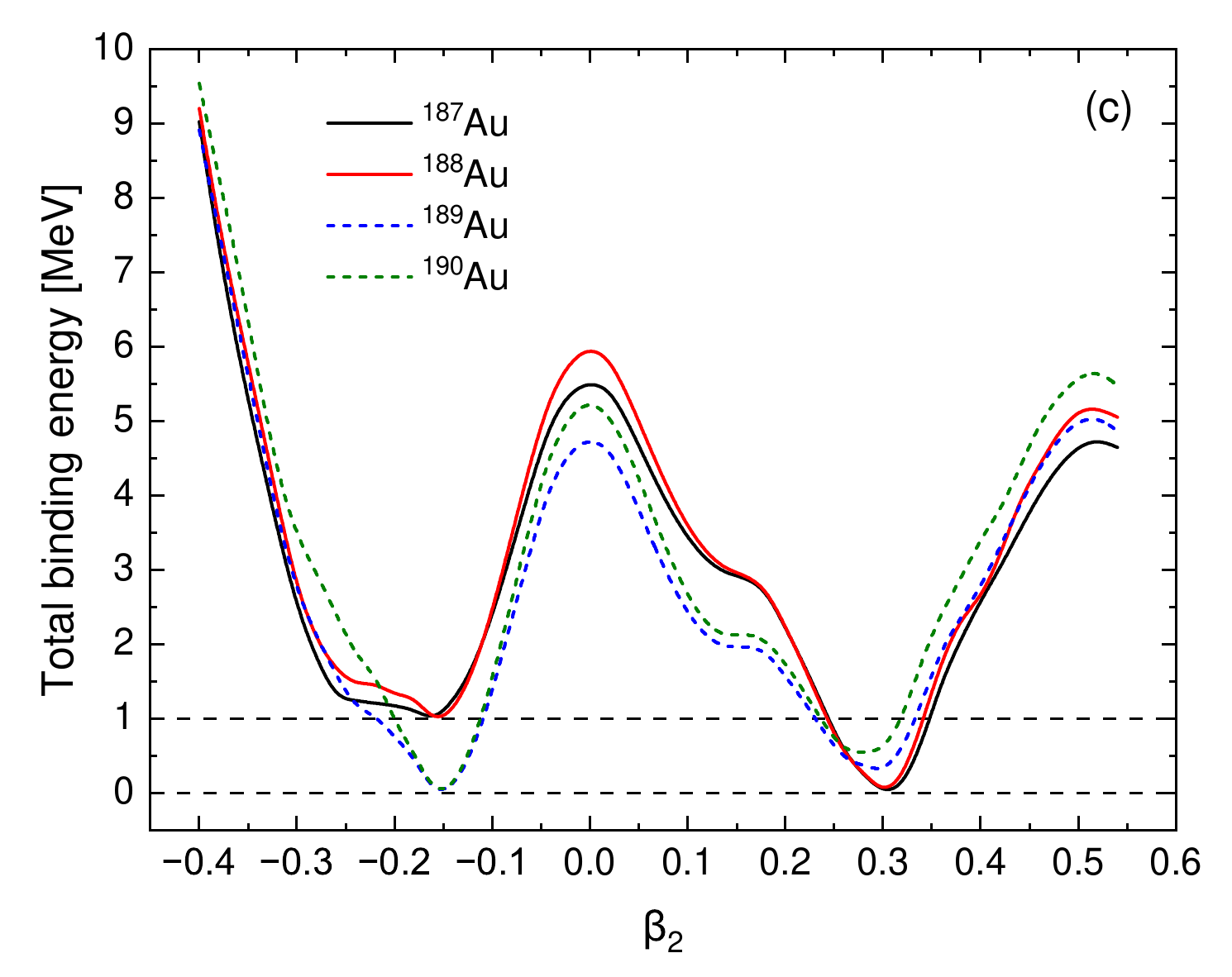}
\includegraphics[width=0.49\linewidth]{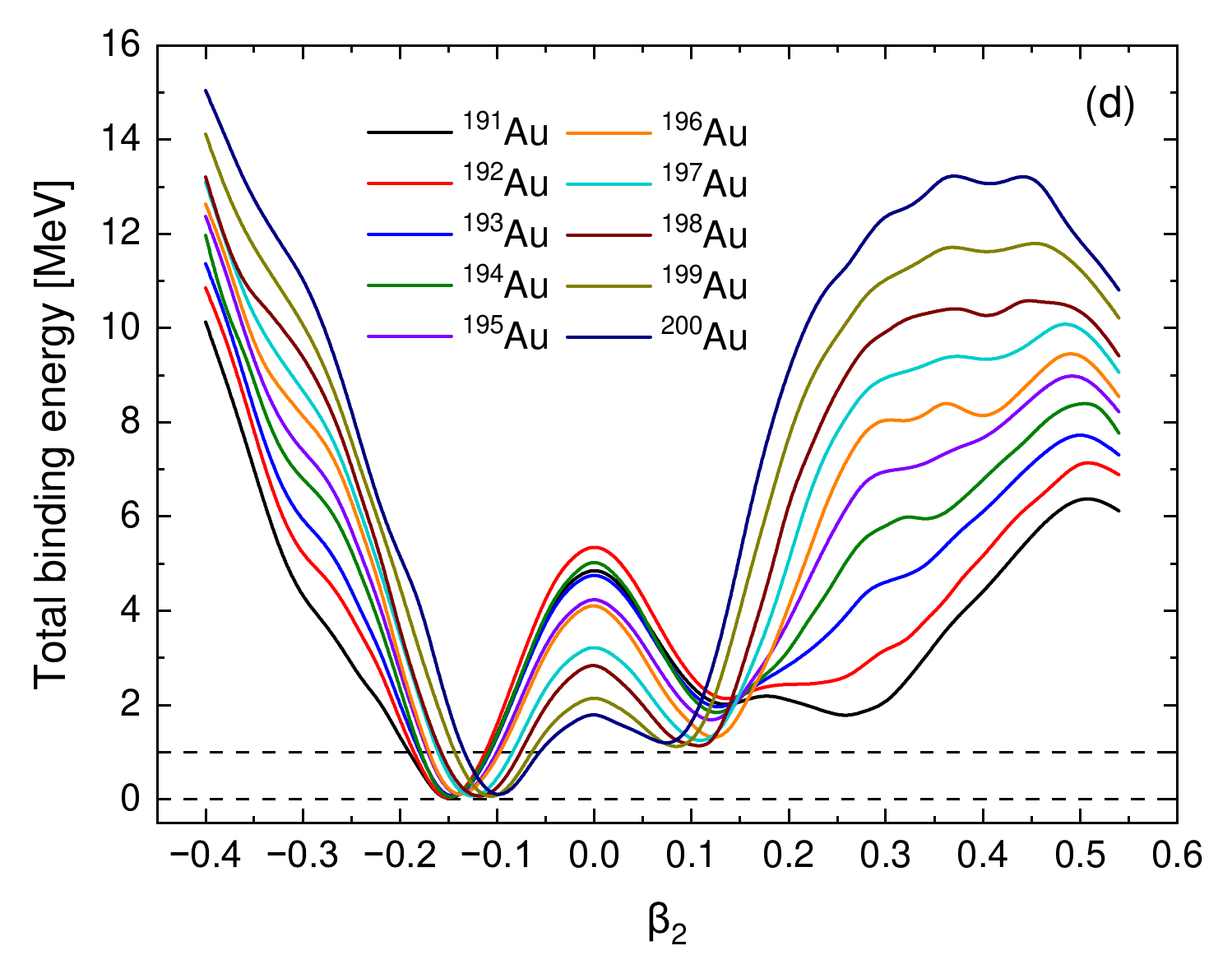}
\caption{(Color online) Evolution of TBE curves in terms of the deformation parameter $\beta_2$ for Au isotopes from the constrained DRHBc calculations. The curves have been scaled to the energy of their ground states. Results of $A = 176 \sim 186$ in the panel (a) and (b) show the prolate minima coresponding to the results by the blue color in Fig. \ref{fig2} (a), while the results of remaind Au isotopes provide oblate deformation. However, $A = 175 \sim 177$ and $A = 179$ in the panel (a) as well as  $A = 187 \sim 190$ in the panel (c) imply shape coexistence possibility.}
\label{fig3}
\end{figure}
 
 \begin{figure}
 \centering
 \includegraphics[width=0.49\linewidth]{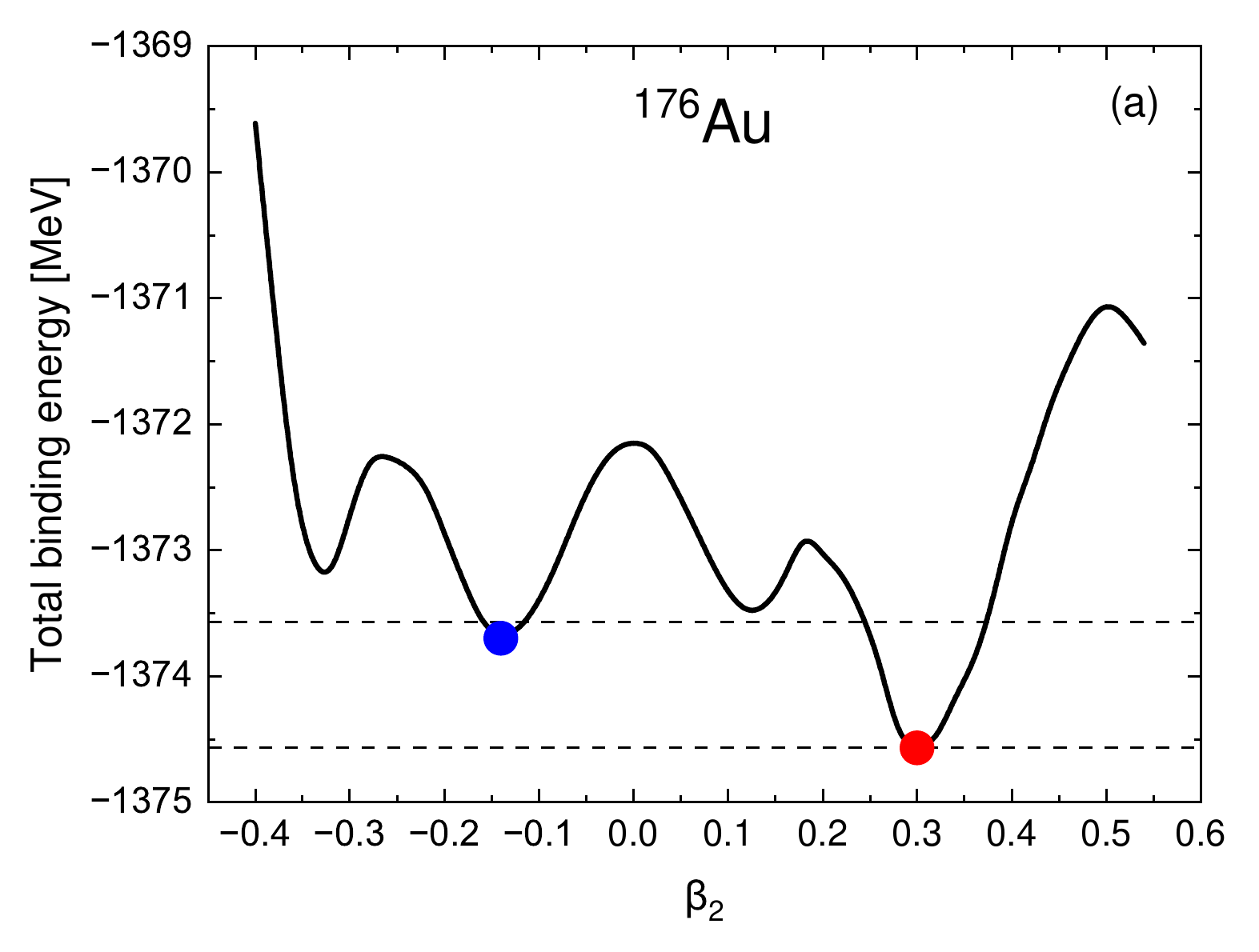}
 \includegraphics[width=0.49\linewidth]{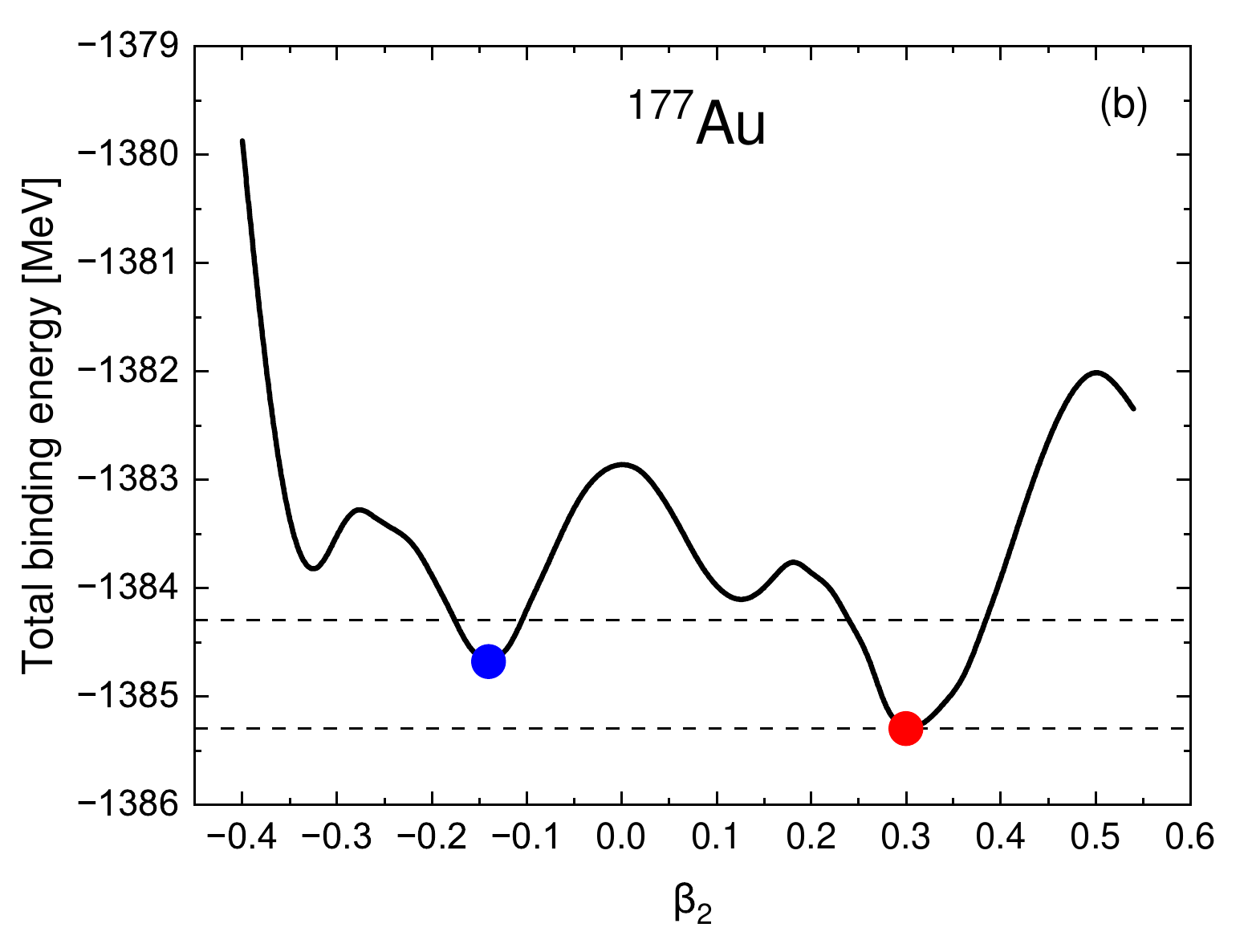}
 \includegraphics[width=0.49\linewidth]{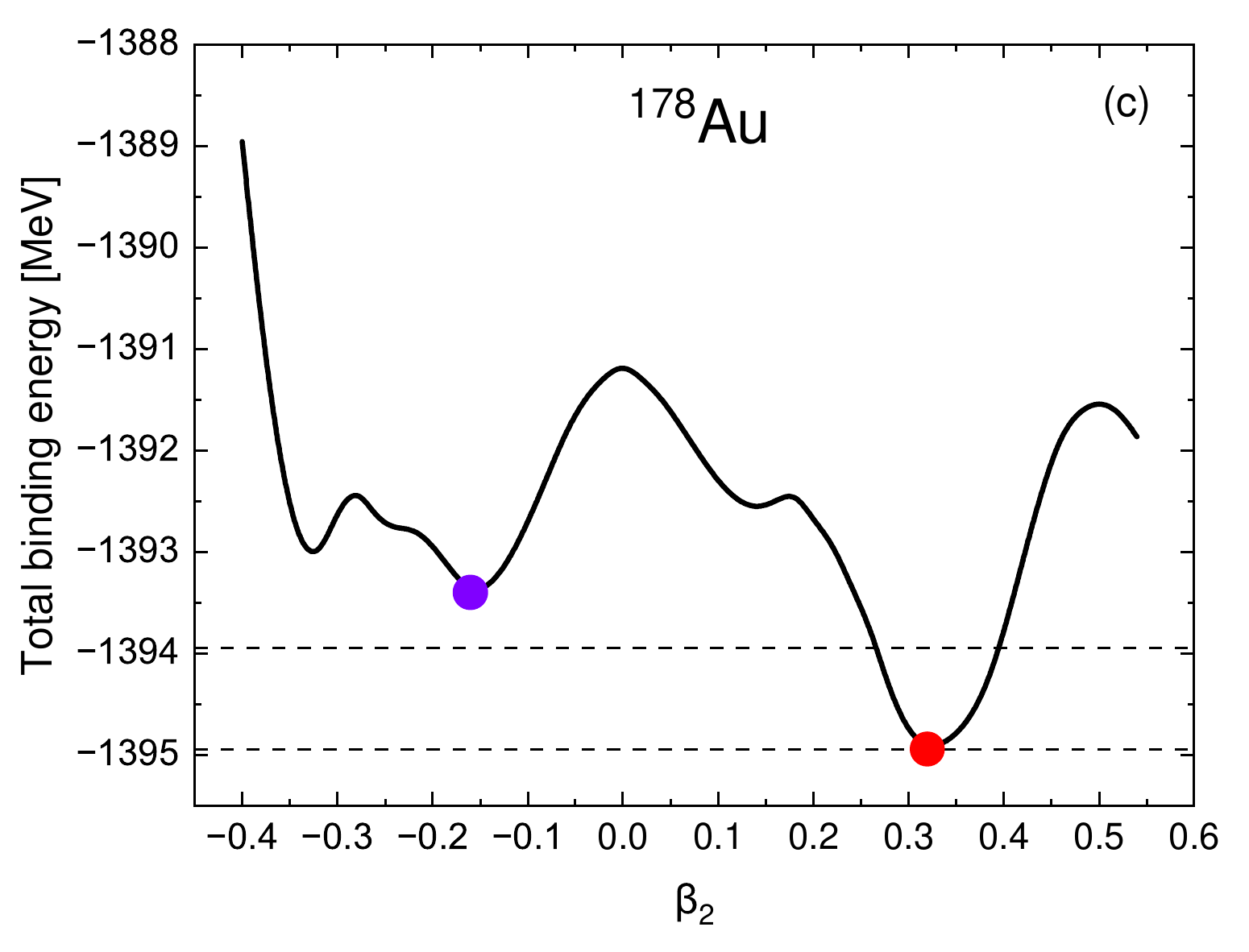}
 \includegraphics[width=0.49\linewidth]{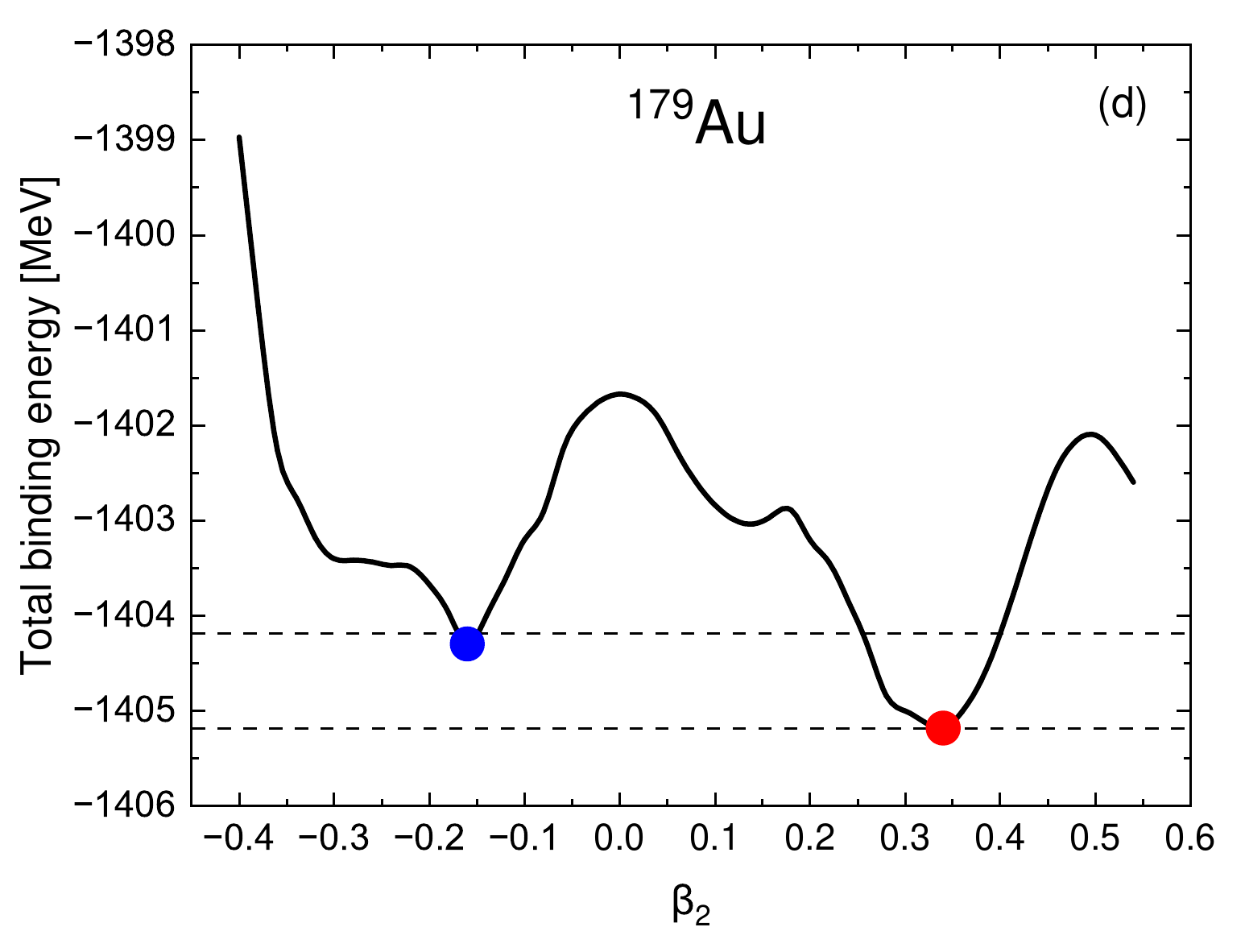}
 \includegraphics[width=0.49\linewidth]{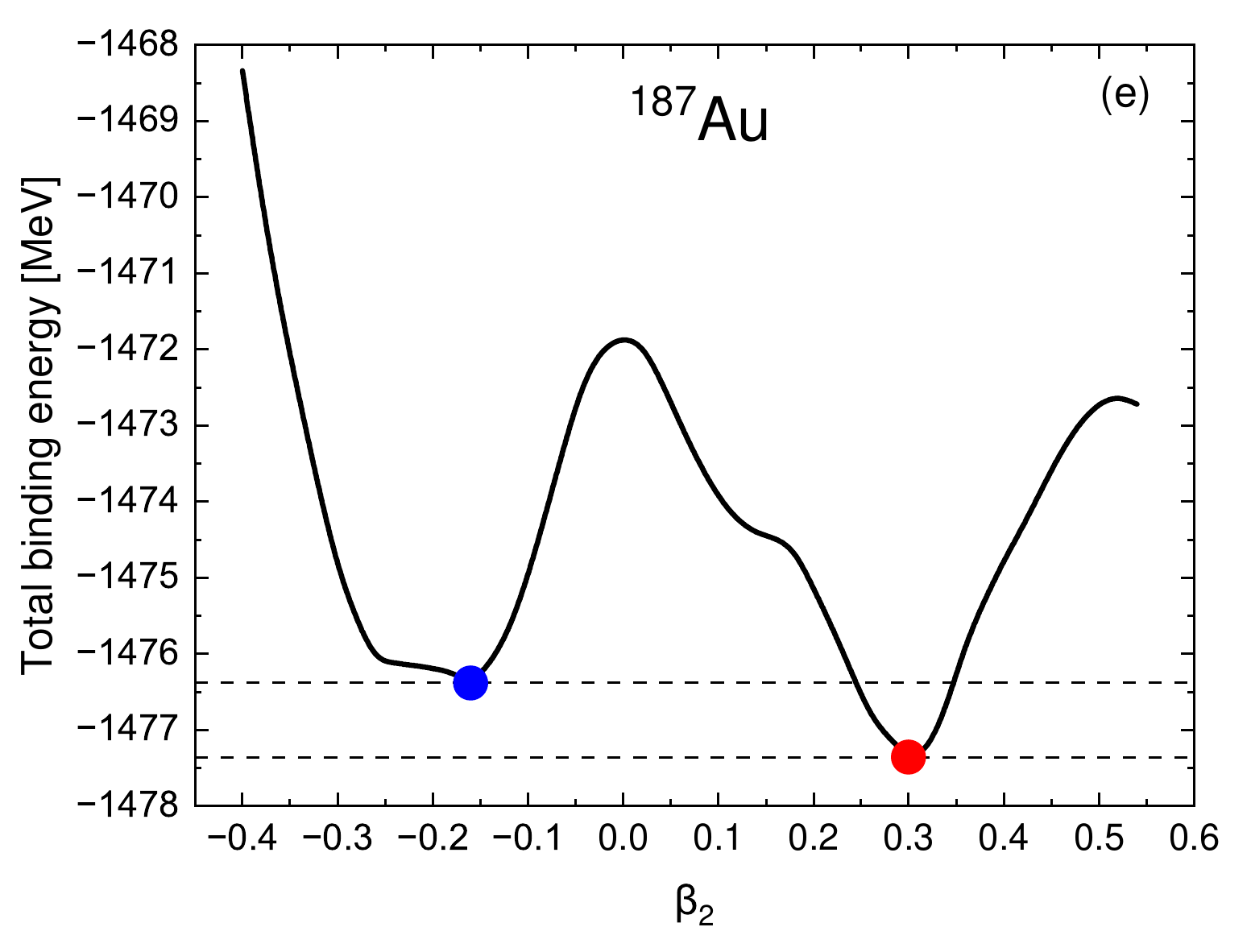}
 \includegraphics[width=0.49\linewidth]{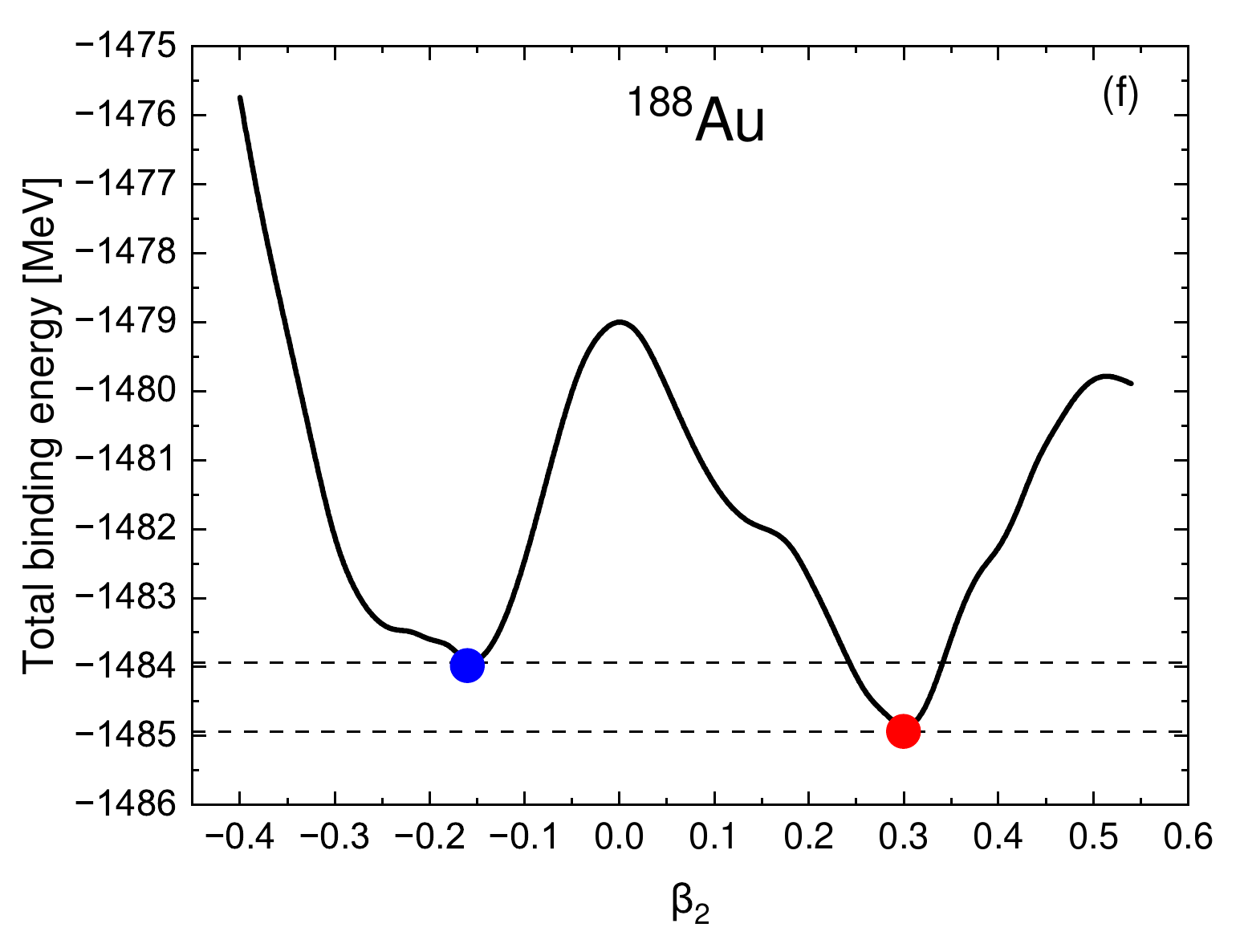}
 \caption{(Color online) Detailed TBE curves in terms of the deformation parameter $\beta_2$ for $^{176-179}$Au and $^{187,188}$Au isotopes. All of the isotopes demonstrate a possibility of the shape coexistence coming from about 1 MeV energy difference between prolate and oblate minima. The violet point in oblate region in the panel (c) for $^{178}$Au discloses local minimum of the TBEs, that is, it is located a bit higher than 1 MeV compared to those in prolate deformation.}
 \label{fig4}
 \end{figure}

\begin{figure}
\centering
\includegraphics[width=0.49\linewidth]{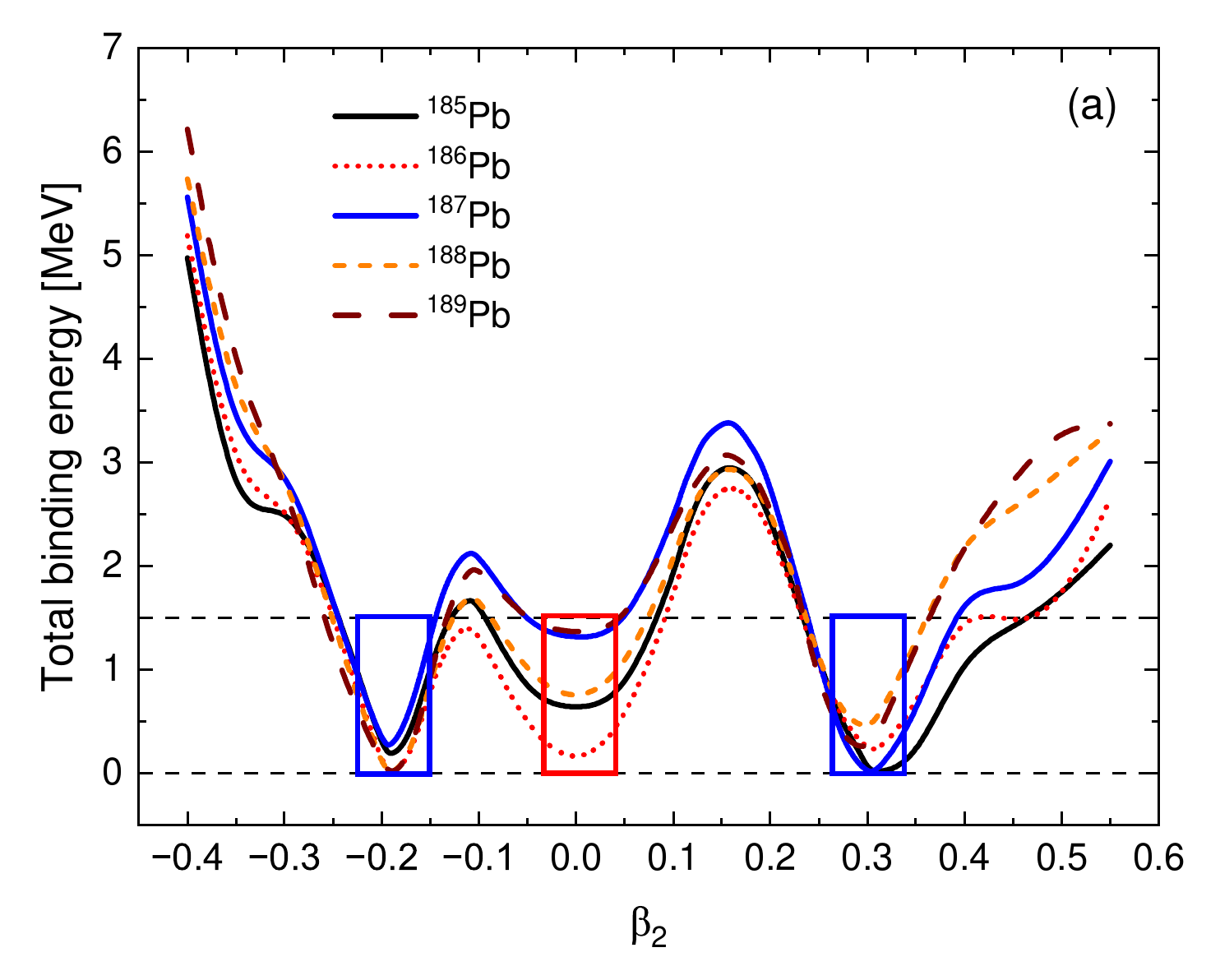}
\includegraphics[width=0.49\linewidth]{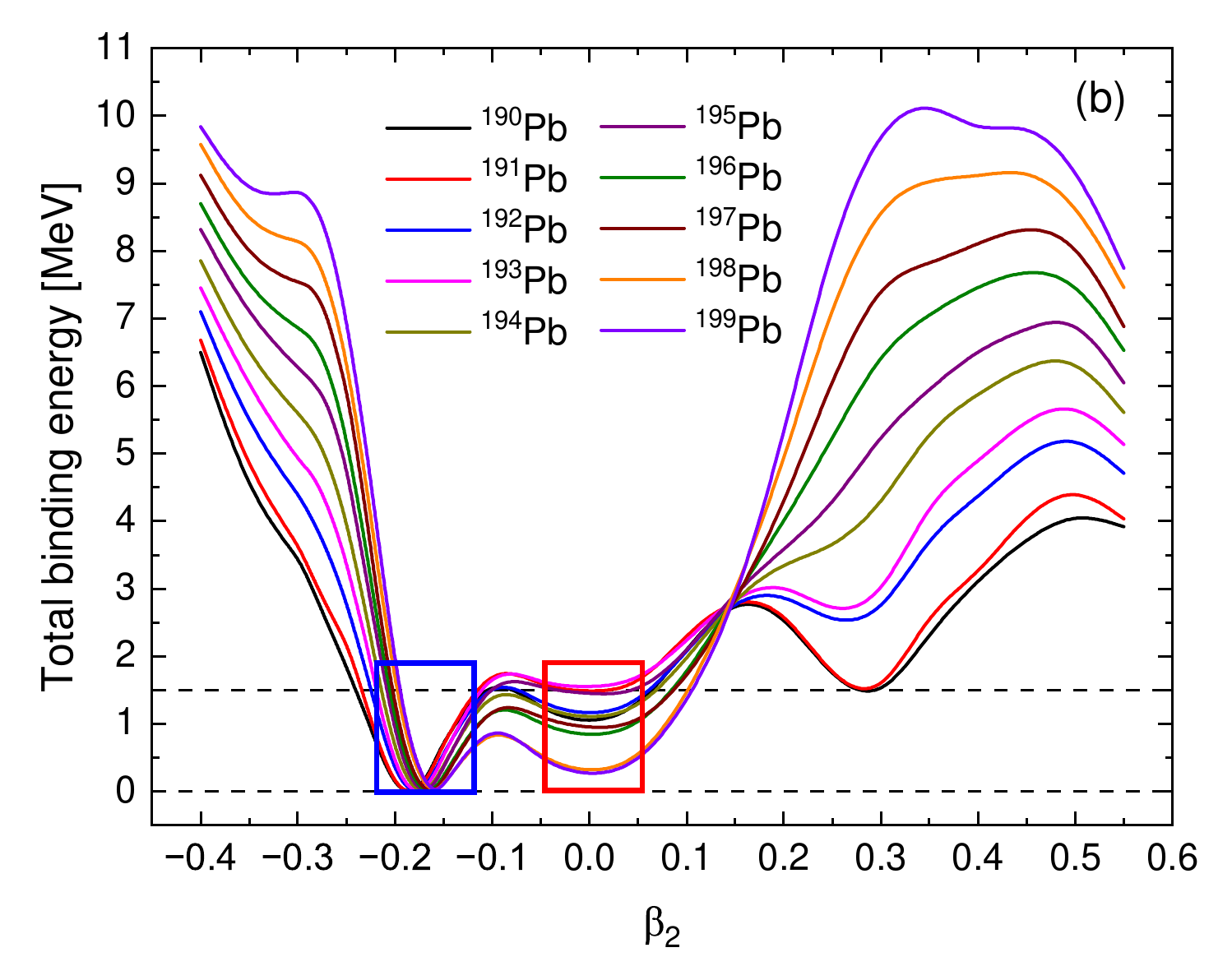}
\caption{(Color online) Evolution of TBE curves in terms of the deformation parameter $\beta_2$ for Pb isotopes from the constrained DRHBc calculations. The curves have been scaled to the energy of their ground states.}
\label{fig5}
\end{figure}
 
 \begin{figure}
 \centering
 \includegraphics[width=0.49\linewidth]{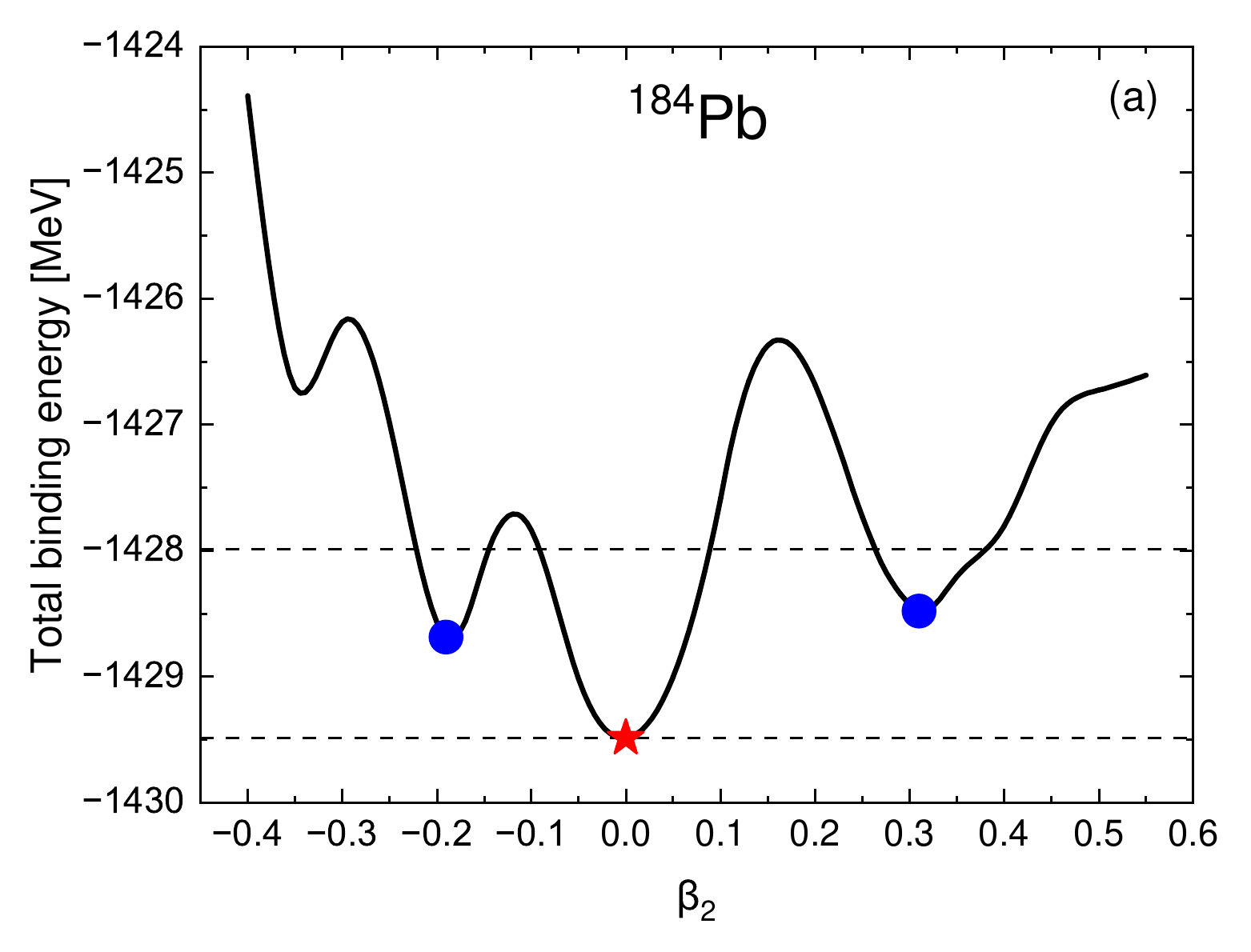}
 \includegraphics[width=0.49\linewidth]{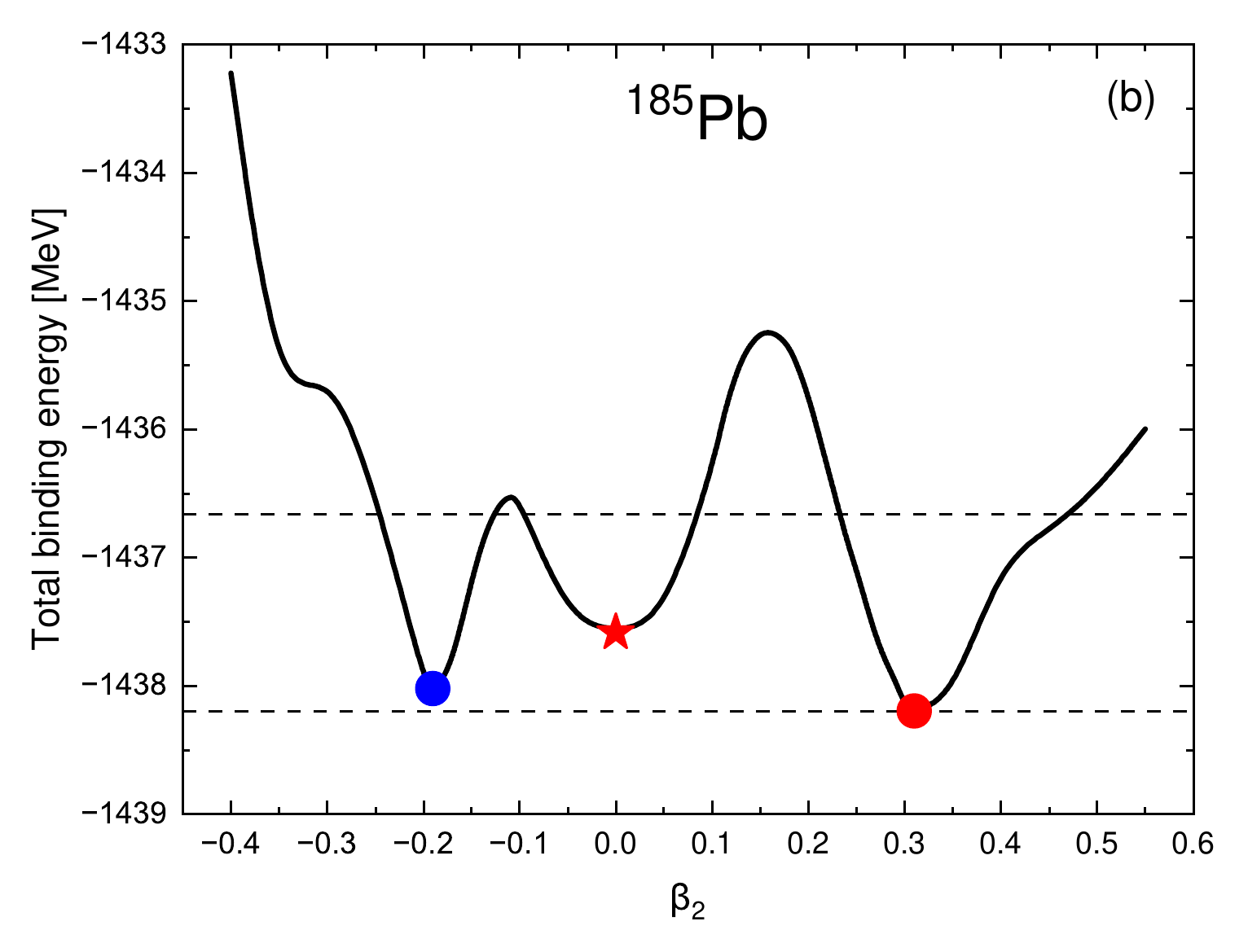}
 \includegraphics[width=0.49\linewidth]{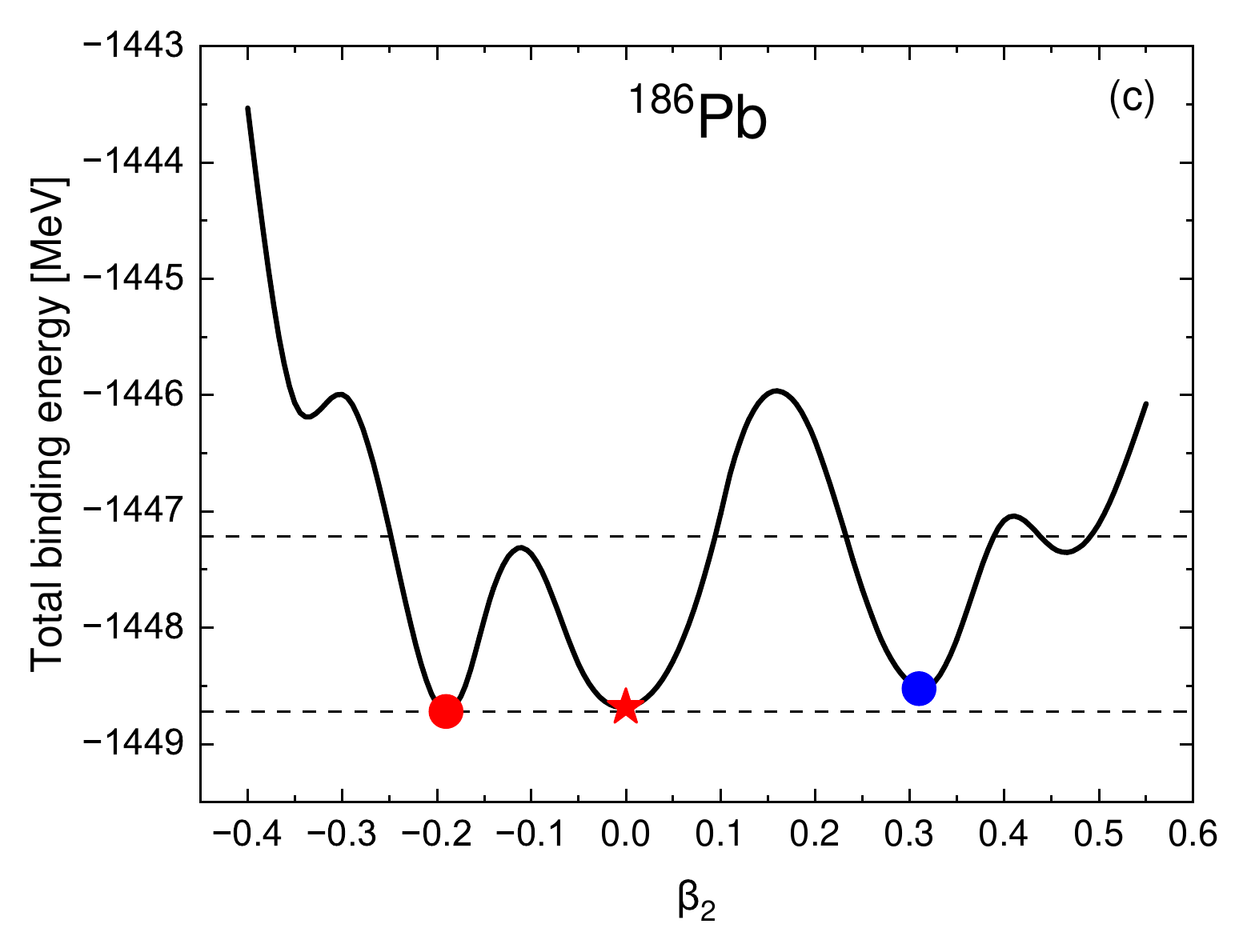}
 \includegraphics[width=0.49\linewidth]{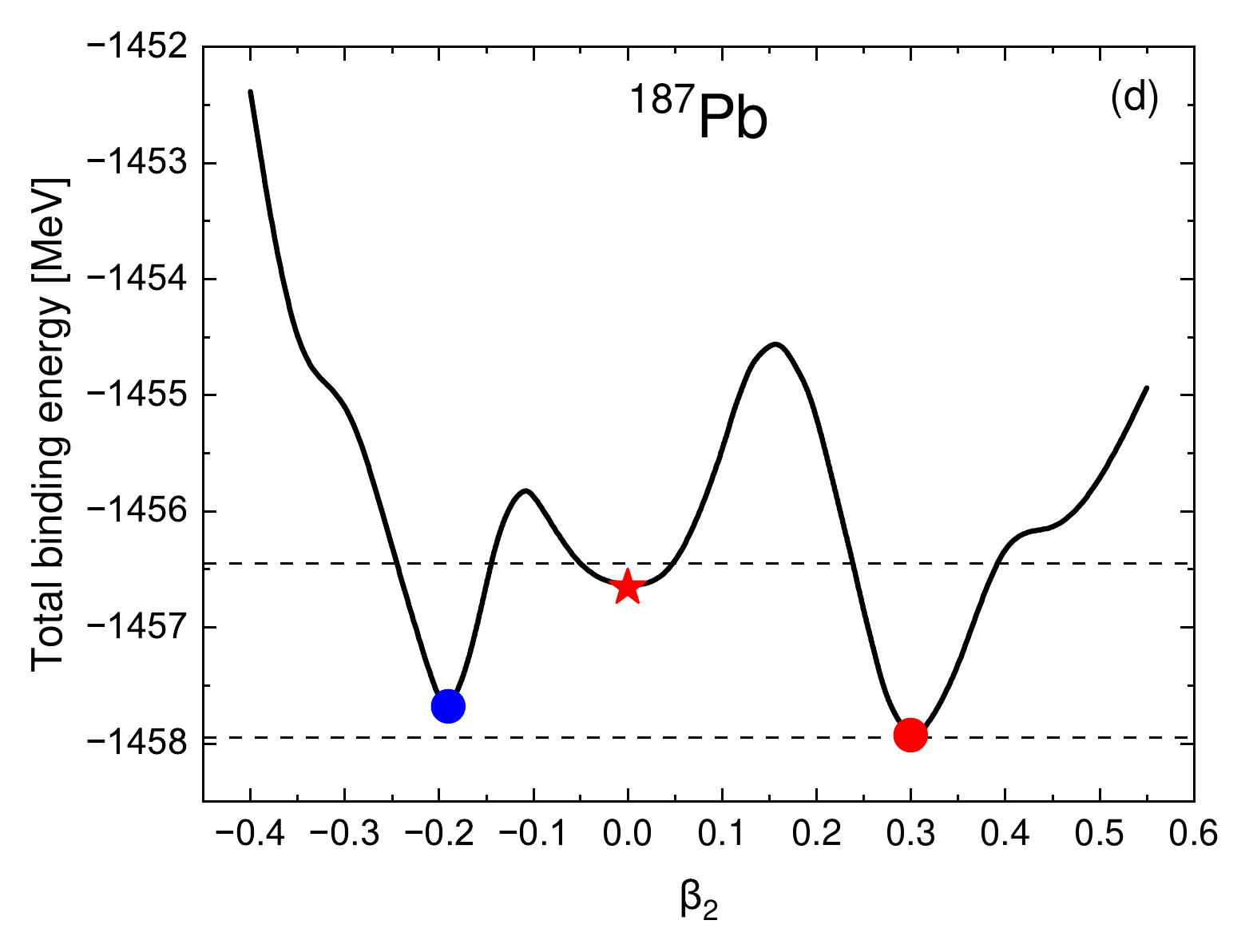}
 \includegraphics[width=0.49\linewidth]{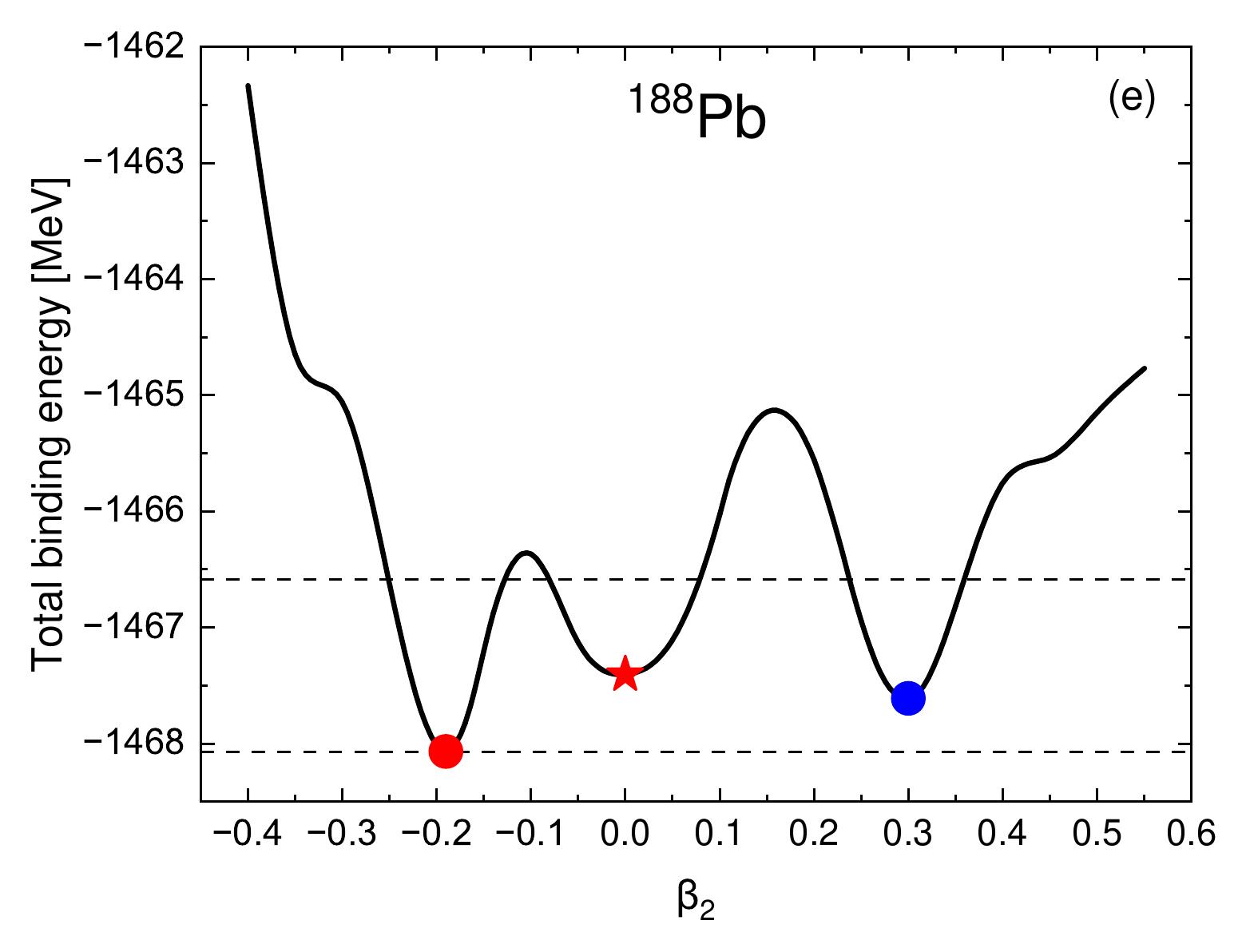}
 \includegraphics[width=0.49\linewidth]{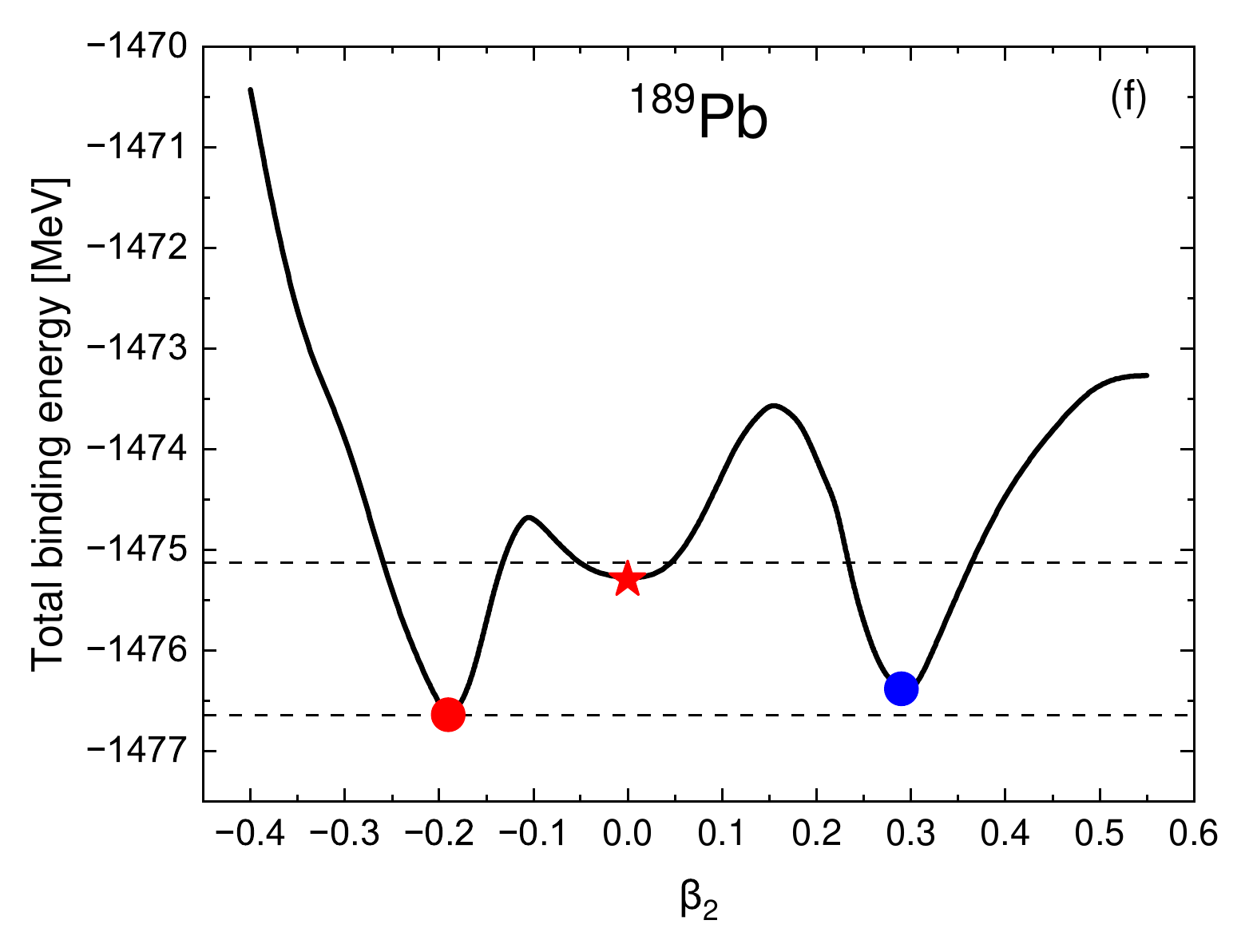}
 \caption{(Color online) Detailed TBE curves in terms of the deformation parameter $\beta_2$ for $^{184-189}$Pb isotopes. All of the isotopes demonstrate a possibility of the shape coexistence coming from about 1.5 MeV energy difference among three deformations.}
 \label{fig6}
 \end{figure}
 
This phenomenon arises from the competition between the prolate minimum and the oblate minimum for Au isotopes (between the oblate minimum and nearly spherical minimum for Pb isotopes). This is clearly depicted in their TBE curves in terms of the deformation parameter $\beta_2$ for Au (Pb) as shown in Fig. \ref{fig3} (Fig. \ref{fig5}). We note the local minima in the prolate (oblate) region in the $175 \leq A \leq 188$ ($A \geq 189$) for Au in Fig. \ref{fig3} and the local minima in the prolate or oblate (nearly spherical) region in the $185 \leq A \leq 199$ ($182 \leq A \leq 184$ and $A \geq 200$) for Pb in Fig. \ref{fig5}, respectively.

To gain a better understanding of the nuclear shape evolution, the detailed TBE curves for Au and Pb isotopes are presented in Fig. \ref{fig4} and \ref{fig6}, respectively. The TBE curves for $^{176-179, 187, 188}$Au and $^{184-189}$Pb in Fig. \ref{fig4} and \ref{fig6} exhibit nearly identical local minima, in two and three deformations, respectively.
This indicates the potential for shape coexistence, which has been extensively discussed in both theoretical nuclear models and experimental studies over the past few decades \cite{Pove2016,Gade2016,Heyde2011,Nach2004,Andr2000,Ojala2022,Paul2009,Kim2022}.
However, it is important to note that, for $^{178}$Au, the difference of TBE values in the oblate deformation (violet circles) is slightly larger than those of other odd and even isotopes (blue circles). This observation offers significant insights for the ensuing discussion on the shape evloution of charge radii in the Au isotopes \cite{Cubiss2023,Sava1990,Pass1994,Blan1997}.

\begin{figure}
\centering
\includegraphics[width=0.49\linewidth]{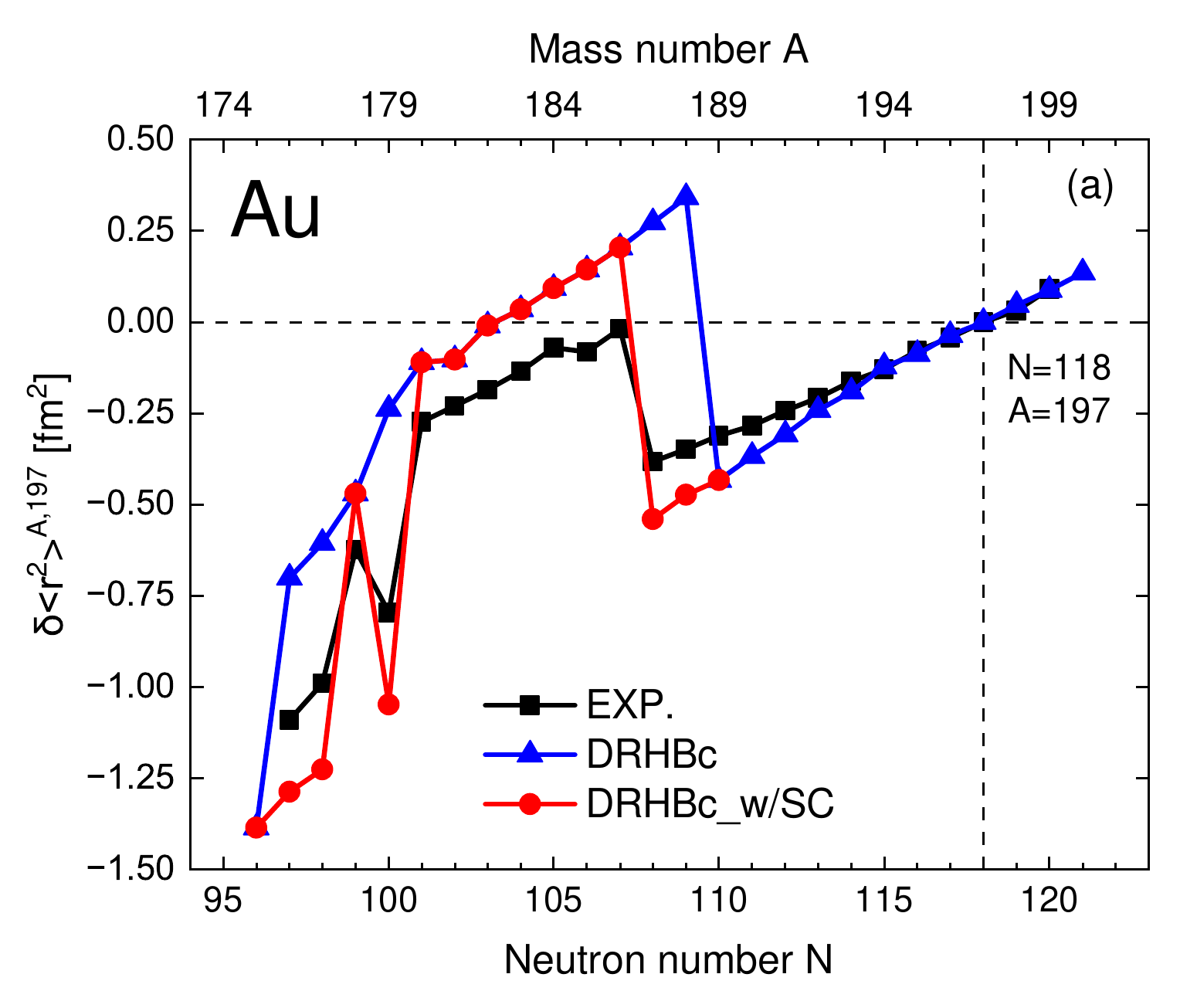}
\includegraphics[width=0.48\linewidth]{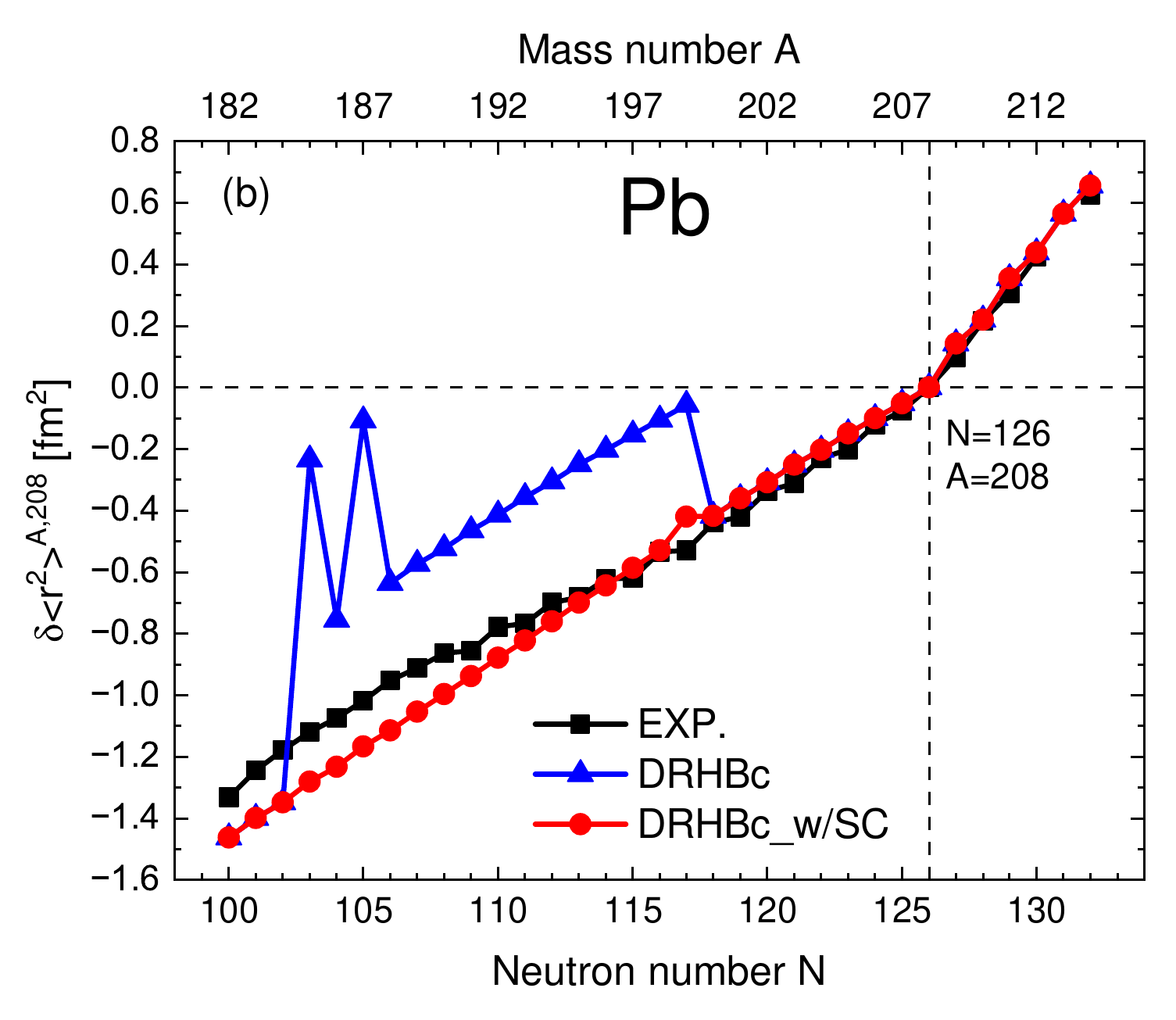}
\caption{(Color online) The relative changes of mean square charge radii $\delta {< r^2 >}^{A,197}$ for Au isotopes with $N = 96 \sim 121$ (a) and $\delta {< r^2 >}^{A,208}$ for Pb isotopes with $N = 100 \sim 132$ (b). Results denoted in a blue color are obtained by prolate shapes calculated by the DRHBc mass model. Red colors in the region are calculated by oblate deformation considering the shape coexistence presented in Fig. \ref{fig3} $\sim$  Fig. \ref{fig6}. Considering the shape coexistence, the odd-even shape staggering of charge radii and the abrupt change of charge rdaii are perspicuously visible for $^{178}$Au isotope and the result of Pb provides a reasonable description. Black boxes for relative changes of mean square charge radii of Au and Pb are taken from experimental data \cite{Cubiss2023,Sava1990,Pass1994,Blan1997,Witte2007,Seli2009}.}
\label{fig7}
\end{figure}

Hereafter, we discuss the shape evloution of charge radii for Au and Pb isotopes. 
Figure \ref{fig7} illustrates the experimental data, which reveal the odd-even shape staggering in the  $N = 98 \sim 100$ region and the sudden change of charge radii above $N =$ 108. Theoretical results are obtained by calculating relative changes in mean square charge radii, $\delta {<r^2>}^{A, A'}~=~<r^2 (A)> - <r^2 (A')>~=~r^2_{ch} (A) - r^2_{ch} (A')$, with respect to the $A'=197$ for Au and $208$ for Pb, respectively.
The results (blue triangles) obtained by prolate (oblate) deformations for $A~\leq~186$ ($A~>~188$) for Au isotopes explain good agreement with the experimental data (black squares), in Fig. \ref{fig7} (a). However, for the $^{176, 177, 179, 187, 188}$Au, the prolate deformation results (blue triangles) overestimate the $\delta {<r^2>}$. Therefore, the blue triangles by prolate deformation of $^{176, 177, 179, 187, 188}$Au isotopes are insufficient in explaining the experimental data. Taking into account the presence of shape coexistence, the results (red circles) obtained using oblate deformation (blue circles) as shown in Fig. \ref{fig4} reasonably describe the relative charge radii data within $\delta {<r^2>} \leq$ 0.25 fm$^2$. Indeed, the abrupt change and the observed shape staggering in the Au isotopes is attributed to the transition from oblate deformation to prolate deformation due to the presence of shape coexistence in the Au isotopes, as seen in Figs. \ref{fig2} and \ref{fig4}. Ultimately, we have successfully reproduced the main features of the experimental results: (i) moving from heavier to lighter masses, the $\delta {<r^2>}$ value increases significantly at $A = 186$ and (ii) the odd-even shape staggering of charge radii is clearly visible for $^{178}$Au isotopes, where the radii for odd-$N$ isotopes are larger than those for neighboring even-$N$ isotopes.

Next, we examine the relative charge radii of Pb isotopes. Figure \ref{fig7} (b) illustrates the evolution of the $\delta {<r^2>}^{A, 208}$ values in the Pb isotopes. In the region where the neutron numbers $N \geq 118$, the relative charge radii of Pb isotopes are reasonably well reproduced. However, for neutron-deficient Pb isotopes within the range of $100 \leq N \leq 117$, the relative charge radii in theoretical results \cite{Pere2021,ADND2022} is somewhat larger  as compared with experiment.

As mentioned in Ref. \cite{ADND2022}, this overestimation can be attributed to their ground states being predicted a large quadrupole deformation, which contradicts experimental observations \cite{Prit2016}. 
Also, upon examination the experimental charge radii in this neutron range maintain the trend at $N = 118 \sim 126$ region (see Figure 7 (a) in Ref. \cite{Pere2021}). In other words, this indicates that the shapes of the nuclei in the measured states are either spherical or near-spherical. Indeed, if we consider spherical solutions in these nuclei, then the experimental data are rather well reproduced.

However, our calculations by DRHBc approach predict either oblate or prolate shapes for the ground states of the $N = 103 \sim 117$ isotopes (see Fig. \ref{fig2} (b)). Despite that, spherical minima exist in all isotopes, either close in energy to the ground states or at some higher energy ($\mid$E$\mid$ $\leq$ 1.5 MeV), as shown in Fig. \ref{fig5} and \ref{fig6}. Therefore, considering shape coexistence similarly to the case of Au, the results (red circles in Fig. {fig7} (b)) for Pb obtained by spherical deformation corresponding to the red square of Fig \ref{fig5} (and the red star in Fig \ref{fig6}) provide a reasonable description of the relative charge radii data within $\delta {<r^2>} \leq$ 0.2 fm$^2$.

\begin{figure}
\centering
\includegraphics[width=0.46\linewidth]{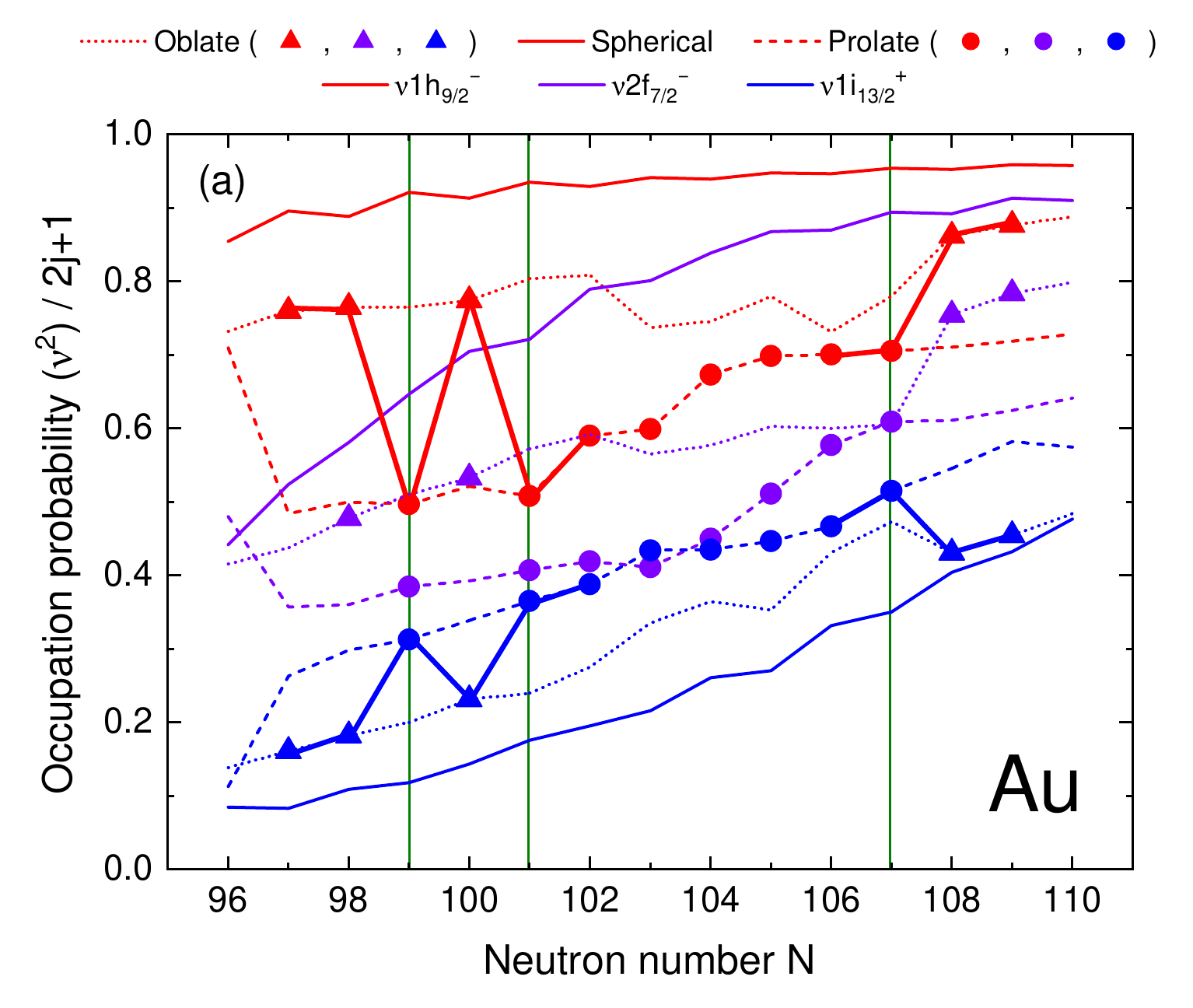}
\includegraphics[width=0.46\linewidth]{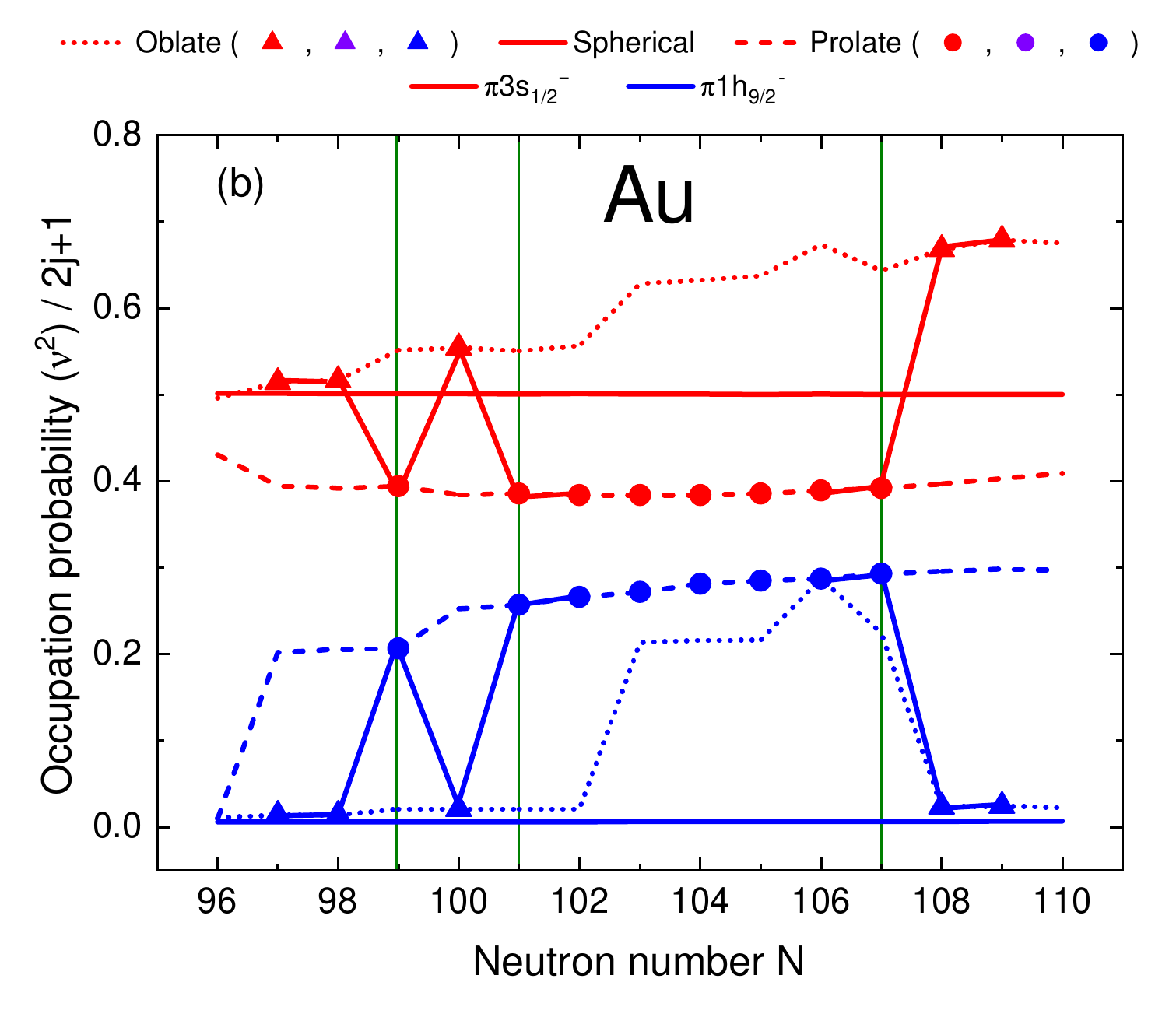}
\includegraphics[width=0.46\linewidth]{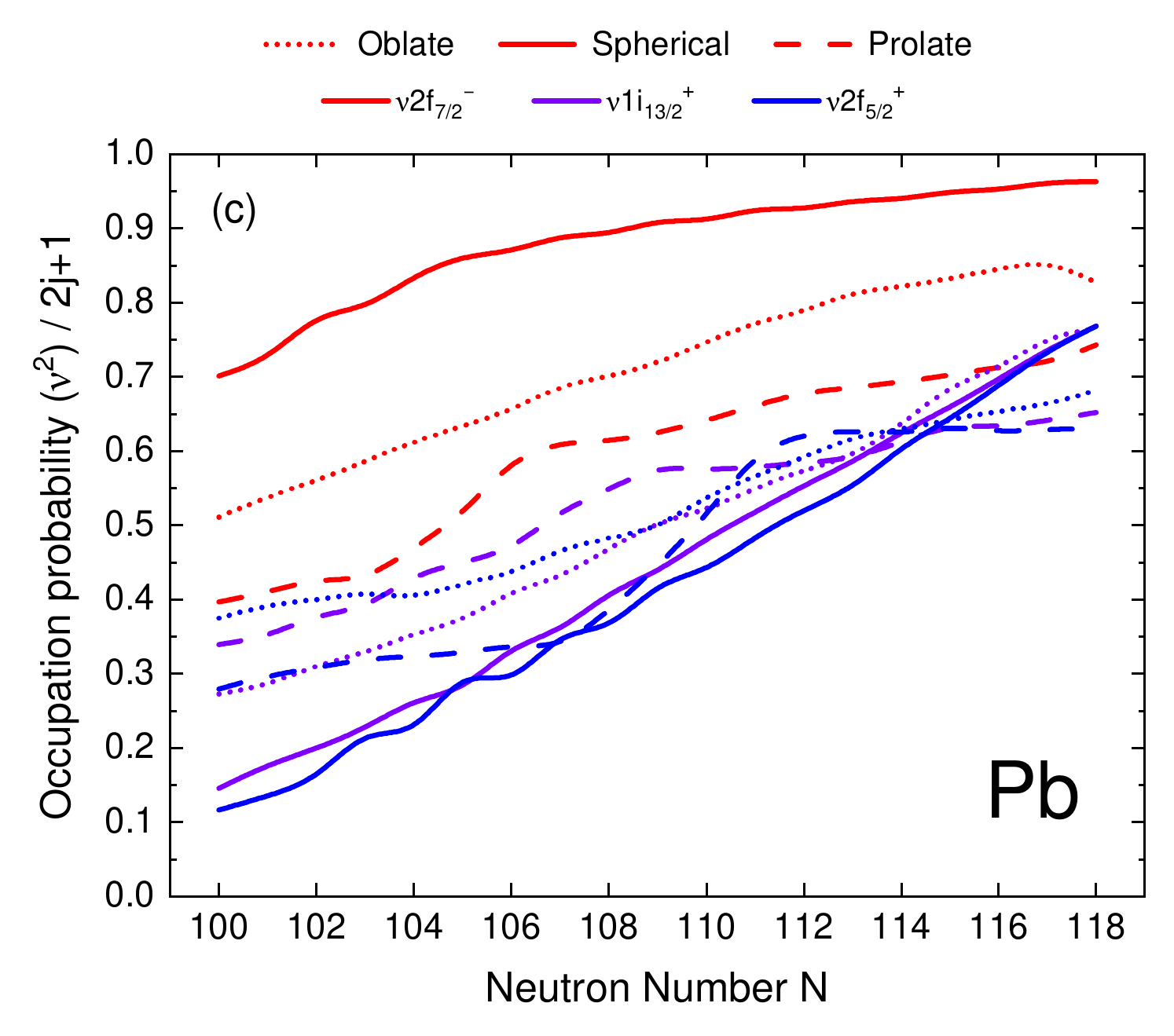}
\includegraphics[width=0.46\linewidth]{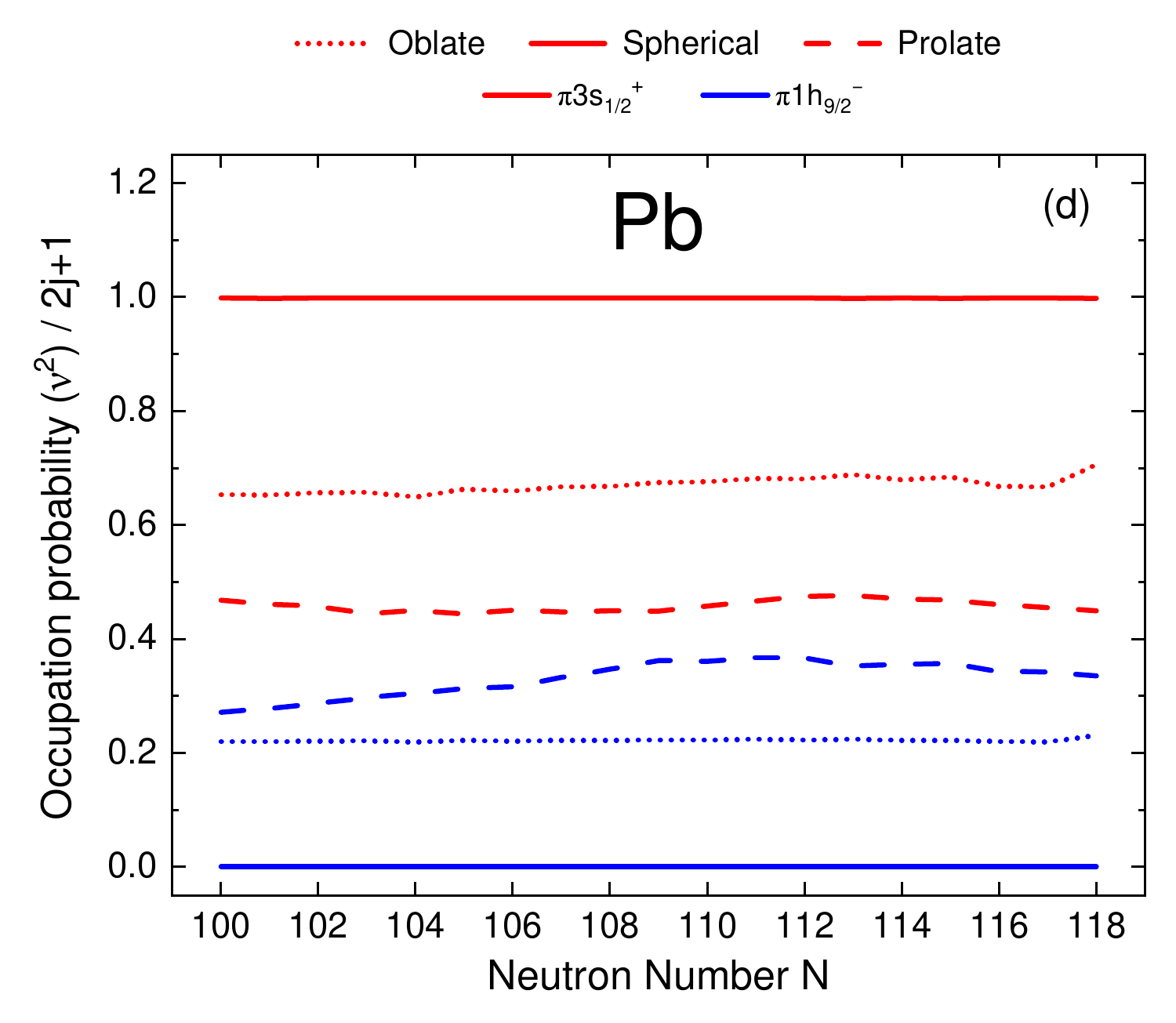}
\caption{(Color online) Evolution of occupation probabilities of the neutron (a) and proton (b) SPSs for the Au isotopes. In prolate deformation region, the occupation probabilities of $\nu 1h_{9/2}$, $\nu 2 f_{7/2}$, and $\pi 3 s_{1/2}$ states decrease while those of ${\nu 1i_{13/2}}$ and $\pi 1 h_{9/2}$ states increase compared to the spherical one. The occupation probabilities of the neutron (c) for Pb isotopes are almost monotonically increased with the increase of neutron number, but those of proton (d) SPSs are unchanged.}
\label{fig8}
\end{figure}

In Fig. \ref{fig8}, the evolution of occupation probabilities for neutron single-particle states, including ${\nu 1 i_{13/2}}$, ${\nu 2 f_{7/2}}$, and ${\nu 1 h_{9/2}}$, as well as proton single-particle states, such as ${\pi 3 s_{1/2}}$ and ${\pi 1 h_{9/2}}$, are depicted for both Au and Pb. We specifically focused on the the occupation probability of ${\nu 1 i_{13/2}}$ state, as noted in Ref. \cite{Mars2018}. The triangles and circles represent the oblate and prolate deformations for Au isotopes, respectively. Remarkably, in Fig. \ref{fig8} (a), one can see that the pronounced increase in the occupation probability of the ${\nu 1 i_{13/2}}$ state explicitly manifests in the prolate deformed $^{178,180,186}$Au isotopes. Conversely, we also observe a decrease in the occupation probability of the ${\nu 1 h_{9/2}}$ and ${\nu 2 f_{7/2}}$ states, which compensates for the abrupt change in the occupation probability of the ${\nu 1 i_{13/2}}$ state. Similarly, in Fig. \ref{fig8} (b), we notice comparable patterns in the proton occupation probabilities for ${\pi 3 s_{1/2}}$ (decrease) and ${\pi 1 h_{9/2}}$ (increase) states compared to the spherical one. As mentioned in the previous paper \cite{Mun2023}, this implicates that the ${\pi 1 h_{9/2}}$ state plays a key role in the abrupt increase of the charge radii of the Au isotopes. These surges can be attributed not only to the quadrupole constituent of the nucleon-nucleon interaction but also to the monopole interaction between ${\nu 1 i_{13/2}}$ state and ${\nu 1 h_{9/2}}$ state \cite{Mars2018}. Meanwhile, the occupancy probabilities of neutrons single-particle states increase, while protons single-particle states for Pb isotopes are unchanged, as shown in Fig. \ref{fig8} (c) and (d). It means that the shape staggering of Au isotopes comes from the shape transitions of neutron and proton single particle state while the increase of charge radii of Pb isotopes are due to the increase of neutron numbers implying no discernible change of symmetric core part composed by $Z = 82$ protons.

The topic of discussion that follows concerns the kink structure, which denotes a rapid increase in charge radii above a magic shell, observed across shell closures \cite{Krei2014}. For example, recent data on Hg isotopes have shown a noticeable kink at $N = 126$ \cite{Good2021,Mars2018}. Moreover, a kink structure has been observed in K and Ca isotopes near the magic shell with $N = 20$ \cite{Kosz2019,Tanaka2020}. In our previous study \cite{Mun2023}, we demonstrated analogous patterns for Hg isotopes at the magic number $N = 126$. Furthermore, we observed  analogous patterns for Pb isotopes beyond a different magic number $N = 184$ \cite{Kim2022}.

\begin{figure}
\centering
\includegraphics[width=0.49\linewidth]{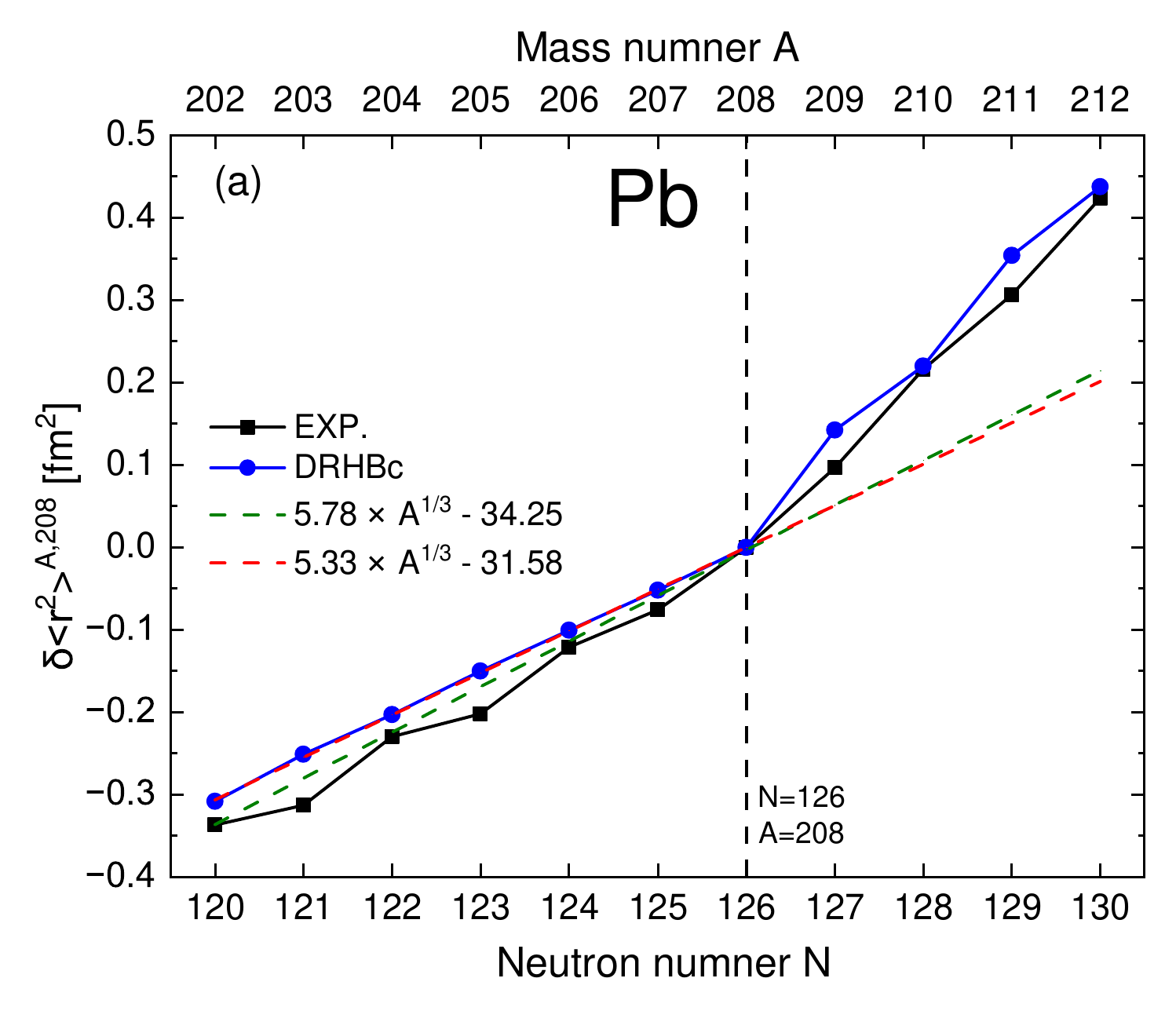}
\includegraphics[width=0.49\linewidth]{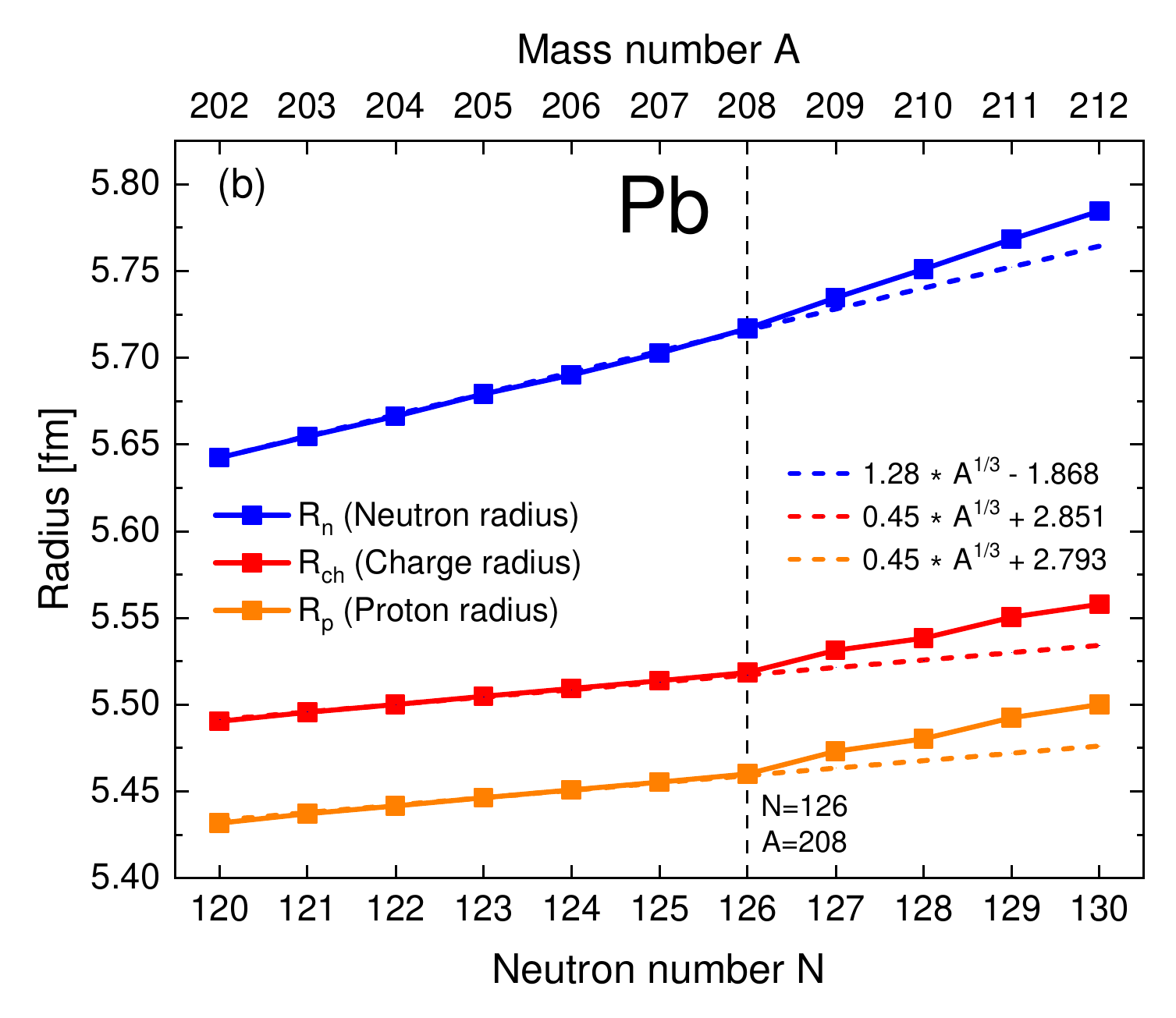}
\includegraphics[width=0.49\linewidth]{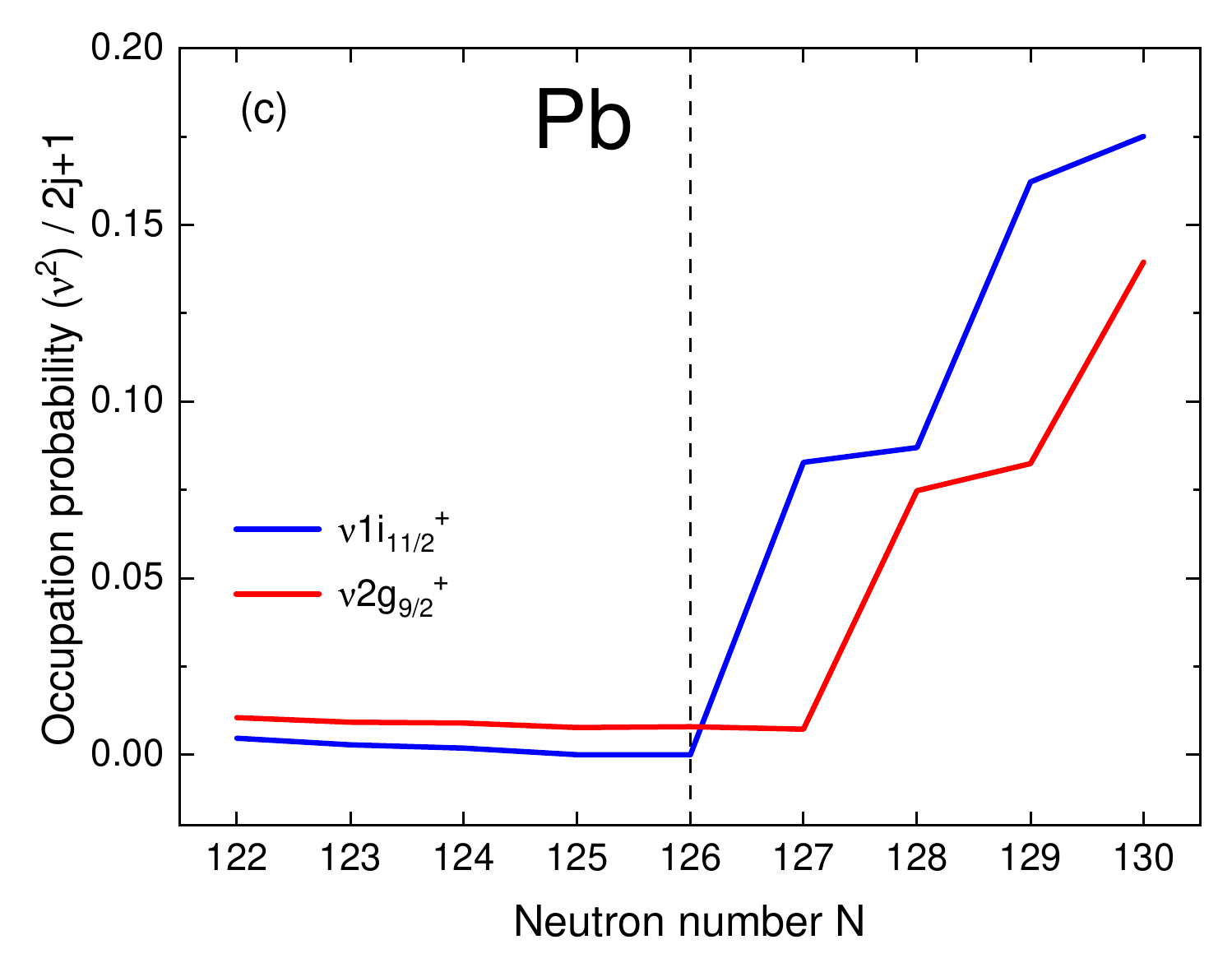}
\includegraphics[width=0.49\linewidth]{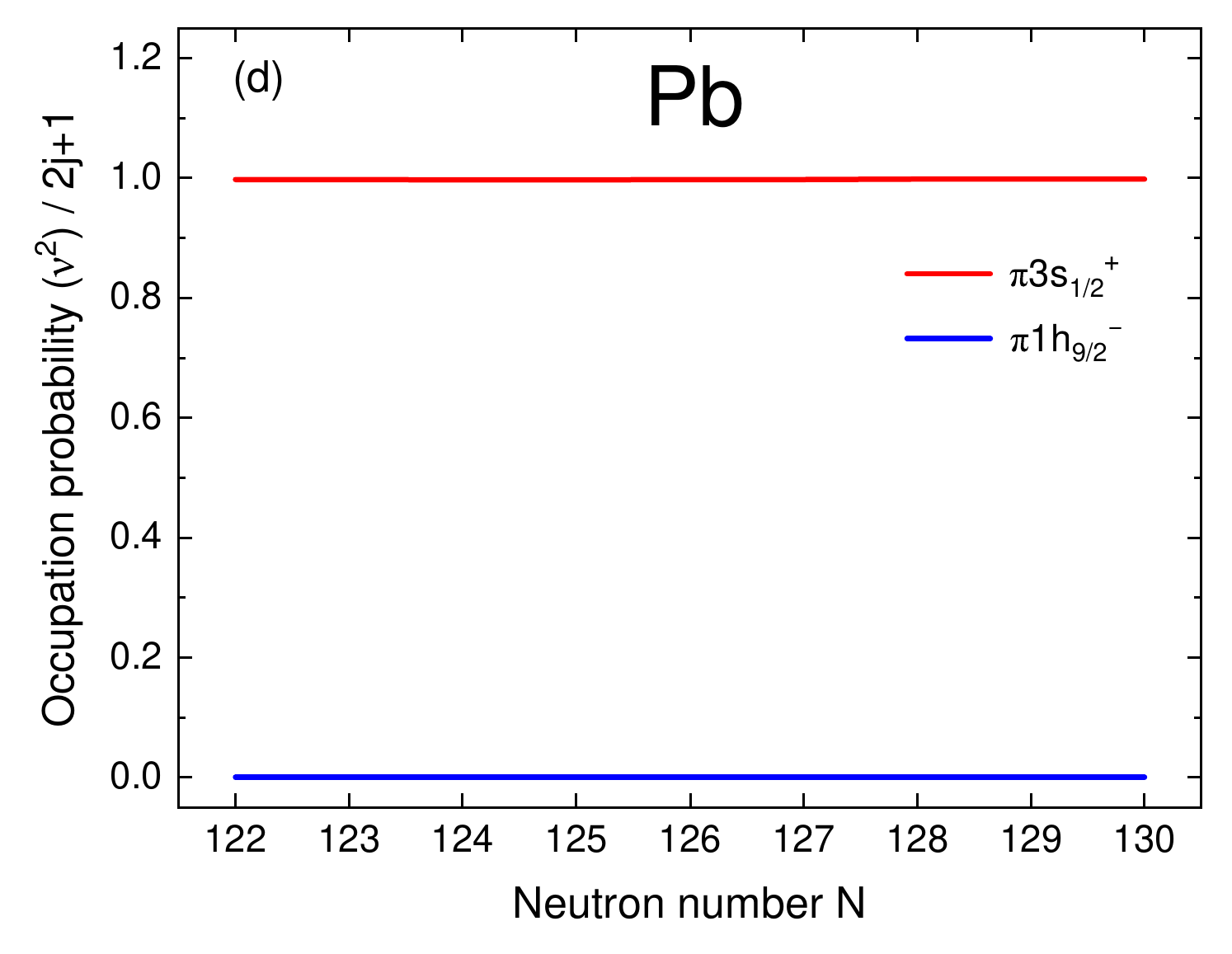}
\caption{(Color online) Kink structure of Pb isotopes in the vicinity of $N =$ 126 for the case normalized at $^{208}$Pb isotope (a) and change of neutron, proton and charge radii nearby $^{208}$Pb. The dashed lines in (a) are just from the continuation of the data before $N =$ 126 case to show the kink structure, while the lines in (b) show how the radii are deviated from the simple continuation. The panel (c) reveals a significant increase in the occupation probabilities of the $\nu 1i_{11/2}$ state compared to the $\nu 2g_{9/2}$ state. In the panel (d), the occupation probabilities of proton states ($\pi 3s_{1/2}$ and $\pi 1h_{9/2}$) are hardly ever changed.}
\label{fig9}
\end{figure}

In Fig. \ref{fig9} (a), our results for Pb isotopes exhibit a kink structure around the closed shell  $N = 126$. Figure \ref{fig9} (b) illustrates the swelling of neutron radii leading to the kink structure from $N = 126$. As mentioned in Refs. \cite{Hori2022,Mun2023}, pairing correlations cause the expansion of the neutron core, leading to an increase of the occupation probabilities for $\nu 2 g_{9/2}$ and $\nu 1 i_{11/2}$ states beyond the $N = 126$ closed shell (Fig. \ref{fig9} (c)). In contrast, in Fig. \ref{fig9} (d), the occupation probabilities of proton states ($\pi 3s_{1/2}$ and $\pi 1h_{9/2}$) are hardly ever changed. This confirms the sense that the kink structure is primarily caused by the swelling of neutrons, particularly of the $\nu 1i_{11/2}$ state, rather than by the increase of the proton $\pi 3s_{1/2}$ and $\pi 1h_{9/2}$ states.

\section{Summary and Conclusion}
 
We found that the shape coexistence is a key property to understand the interesting nuclear shape evolution. If we consider the shape coexistence of certain Au isotopes, we can certainly demonstrate two phenomena: (i) as we move from heavier to lighter masses, the $\delta {<r^2>}$ value increases significantly at $A = 186$, and (ii) the odd-even shape staggering of charge radii is perspicuously visible for $^{178}$Au. 
Here, we adopt the oblate deformation of $^{176, 177, 179, 187, 188}$Au because their TBEs by oblate deformation are about 1 MeV smaller than those by prolate deformation. However, other nuclei (up to $^{186}$Au) exhibit prolate shapes, because their local minima of TBEs for oblate deformation are 1 MeV larger than the minima for prolate deformation. Therefore, these prolate nuclei, which are excluded from the shape coexistence, do not exhibit significant change of charge radii. This implies that the odd-even shape staggering and the abrupt change could be a strong indicator of the shape coexistence of some Au isotopes, or vice versa. Here we note that possible existence of triaxial deformation may lead the minima calculated in the present work to a saddle point in the $(\beta, \gamma)$ plane \cite{YLYang2021}. We leave the possibility of the $\gamma$ deformation in these isotopes as a future work.

Furthermore, within the DRHBc approach, we have successfully reproduced the kink structure of Pb observed around the $N = 126$ shell. This kink structure is caused by the increase in the occupation probabilities of $\nu 2 g_{9/2}$ and $\nu 1 i_{11/2}$ states beyond the closed shell at $N = 126$. The observed kink structure is closely linked to the swelling of the symmetric core at $N = 126$ followed by the pulling of the protons by the symmetry energy. We intend to further investigate the significant nuclear shape evolution and the kink structure of charge radii in K, Ca, and Bi isotopes near each magic shell using the DRHBc model in future studies.

\section{Acknowledgments}
Helpful discussions with members of the DRHBc Mass Table Collaboration are gratefully appreciated.
This work was supported by the National Research Foundation of Korea (NRF) grant funded by the Korea government (Grant Nos. NRF-2021R1A6A1A03043957, NRF-2020R1A2C3006177, NRF-2021R1F1A1060066, NRF-2018R1D1A1B05048026). 
This work was supported by the National Supercomputing Center with supercomputing resources including technical support (KSC-2023-CRE-0521 and TS-2024-RE-0024)


\begin{thebibliography}{99}
\bibitem{Angeli2013} I. Angeli, K. P. Marinova, At. Data Nucl. Data Tables {\bf 99}, 69 (2013).
\bibitem{Angeli2015} I. Angeli, K. P. Marinova, J. Phys. G: Nucl. Part. Phys. {\bf 42}, 055108 (2015).
\bibitem{Wang2013} Ning Wang and Tao Li, Phys. Rev. C {\bf 88}, 011301(R) (2013).
\bibitem{Gorges2019} C. Gorges, L. V. Rodr\'{i}guez, D. L. Balabanski, M. L. Bissell, K. Blaum, B. Cheal {\it et al.}, Phys. Rev. Lett. {\bf 122}, 192502 (2019).
\bibitem{Wood1992} J. L. Wood, K. Heyde, W. Nazarewicz, M. Huyse and P. van Duppen, Phys. Rep. {\bf 215}, 101 (1992).
\bibitem{Cejnar2010} P. Cejnar, J. Jolie, R. F. Casten, Rev. Mod. Phys. {\bf 82}, 2155 (2010).
\bibitem{Witte2007} H. De Witte, A. N. Andreyev, N. Barr\'{e} M. Bender, T. E. Cocolios, S. Dean {\it et al.}, Phys. Rev. Lett. {\bf 98}, 112502 (2007)
\bibitem{Flanagan2013} K. T. Flanagan, K. M. Lynch, J. Billowes, M. L. Bissell, I. Budinčević, T. E. Cocolios {\it et al.}, Phys. Rev. Lett. {\bf 111}, 212501 (2013).
\bibitem{Bao2021} M. Bao and Q. Wei, Symmetry {\bf 13}, 2253 (2021).
\bibitem{Geithner2008} W. Geithner, T. Neff, G. Audi, K. Blaum, P. Delahaye, H. Feldmeier {\it et al.}, Phys. Rev. Lett. {\bf 101}, 252502  (2008).
\bibitem{Nort2009} W. Nörtershäuser, D. Tiedemann, M. Žáková, Z. Andjelkovic, K. Blaum, M. L. Bissell {\it et al.}, Phys. Rev. Lett. {\bf 102}, 062503 (2009).
\bibitem{Yordanov2016} D. T. Yordanov, D. L. Balabanski, M. L. Bissell, K. Blaum, I. Budinčević, B. Cheal {\it et al.}, Phys. Rev. Lett. {\bf 116}, 032501 (2016).
\bibitem{Bohr} A. Bohr and B. R. Mottelson, \textit{Nuclear Structure, Vol. II} (W. A. Benjamin Inc., 1975; World Scientific, 1998).
\bibitem{Duflo1994} J. Duflo, Nucl. Phys. A {\bf 576}, 29 (1994).
\bibitem{Die2009} A. E. L. Dieperink and P. Van Isacker, Eur. Phys. J. A {\bf 42}, 269 (2009).
\bibitem{Qian2014} Y. B. Qian, Z. Z. Ren, and D. D. Ni, Phys. Rev. C {\bf 89}, 024318 (2014).
\bibitem{Ma2017} W. H. Ma, J. S. Wang, S. Mukherjee, Q. Wang,  D. Patel, Y.-Y Yang {\it et al.}, Chin. Phys. C {\bf 41}, 044103 (2017).
\bibitem{Li2021} T. Li, Y. Luo and N. Wang, At. Data Nucl. Data Tables {\bf 140}, 101440 (2021).
\bibitem{Vries1987} H. De Vries, C. De Jager, and C. De Vries, At. Data Nucl. Data Tables {\bf 36}, 495 (1987).
\bibitem{Avgo2011} M. Avgoulea, Y. P. Gangrsky, K. P. Marinova, S. G. Zemlyanoi, S. Fritzsche, D. Iablonskyi {\it et al.}, J. Phys. G {\bf 38}, 025104 (2011).
\bibitem{Mari2015} K. Marinova, J. Phys. Chem. Ref. Data {\bf 44}, 031214 (2015).
\bibitem{Angeli2016} I. Angeli, K. Marinova, J. Phys. Conf. Ser. {\bf 724}, 012032 (2016).
\bibitem{Boehm1974} F. Boehm and P. L. Lee, At. Data Nucl. Data Tables {\bf 14}, 605 (1974). 
\bibitem{Mano2019} T. Manovitz, R. Shaniv, Y.Shapira, R. Ozeri, and N. Akerman, Phys. Rev. Lett. {\bf 123}, 203001 (2019).
\bibitem{Han2022} J. Z. Han, C. Pan, K. Y. Zhang, X. F. Yang, S. Q. Zhang, J. C. Berengut {\it et al.}, Phys. Rev. Research {\bf 4}, 033049 (2022).
\bibitem{Krie2012} A. Krieger, K. Blaum, M.L. Bissell {\it et al.}, Phys. Rev. Lett. {\bf 108} 142501 (2012).
\bibitem{Seli2013} M. D. Seliverstov, T. E. Cocolios, W. Dexters, A. N. Andreyev, S. Antalic, A. E. Barzakh {\it et al.}, Phys. Lett. B {\bf 719}, 362 (2013).
\bibitem{Mina2016} K. Minamisono, D.M. Rossi, R. Beerwerth {\it et al.}, Phys. Rev. Lett. {\bf 117}, 252501 (2016).
\bibitem{Marsh2018} B. A. Marsh, T. Day Goodacre, S. Sels {\it et al.}, Nat. Phys. {\bf 14}, 1163 (2018).
\bibitem{Kauf2020} S. Kaufmann, J. Simonis, S. Bacca {\it et al.}, Phys. Rev. Lett. {\bf 124}, 132502 (2020).
\bibitem{Miller2019} A. J. Miller, K. Minamisono, A. Klose {\it et al.}, Nat. Phys. {\bf 15}, 432 (2019).
\bibitem{Garc2016} R. F. Garcia Ruiz, M.L. Bissell, K. Blaum {\it et al.}, Nat. Phys. {\bf 12}, 594 (2016).
\bibitem{Groote2020} R. P. de Groote, J. Billowes, C.L. Binnersley {\it et al.}, Nat. Phys. {\bf 16}, 620 (2020).
\bibitem{Cubiss2023} J. G. Cubis, A. N. Andreyev, A. E. Barzakh, P. Van Duppen, S. Hilaire, S. P\'{e}ru {\it et al.}, Phys. Rev. Lett. {\bf 131}, 202501 (2023).
\bibitem{Rose2015} M. Rosenbusch, P. Ascher, D. Atanasov, C. Barbieri, D. Beck, K. Blaum {\it et al.}, Phys. Rev. Lett. {\bf 114}, 202501 (2015). 
\bibitem{Krei2014} K. Kreim, M. Bissell, J. Papuga, K. Blaum, M. De Rydt, R. F. Garcia Ruiz {\it et al.}, Phys. Lett. B {\bf 731}, 97 (2014).
\bibitem{Kosz2019} \'{A}. Koszor\'{u}s, X. F. Yang, J. Billowes, C. L. Binnersley, M. L. Bissell, T. E. Cocolios {\it et al.}, Phys. Rev. C {\bf 100}, 034304 (2019).
\bibitem{Kosz2021} \'{A}. Koszor\'{u}s, X. F. Yang, W. G. Jiang, S. J. Novario, S. W. Bai, J. Billowes {\it et al.}, Nature Phys. {\bf 17}, 439 (2021), [Erratum: Nature Phys. {\bf 17}, 539 (2021)].
\bibitem{Geld2022} S. Geldhof, M. Kortelainen, O. Beliuskina, P. Campbell, L. Caceres, L. Ca\~{n}ete {\it et al.}, Phys. Rev. Lett. {\bf 128}, 152501 (2022). 
\bibitem{Good2021} T. Day Goodacre, A. V. Afanasjev, A. E. Barzakh, B. A. Marsh, S. Sels, P. Ring {\it et al.}, Phys. Rev. Lett. {\bf 126}, 032502 (2021).
\bibitem{Tanaka2020} M. Tanaka, M. Takechi, M. Fukuda, D. Nishimura, T. Suzuki, Y. Tanaka {\it et al.}, Phys. Rev. Lett. {\bf 124}, 102501 (2020). 
\bibitem{Wien2013} F. Wienholtz, D. Beck, K. Blaum, Ch. Borgmann, M. Breitenfeldt, R. B. Cakirli {\it et al.}, Nature {\bf 498}, 346 (2013). 
\bibitem{Gorg2019} C. Gorges, L. V. Rodr\'{i}guez, D. L. Balabanski, M. L. Bissell, K. Blaum, B. Cheal {\it et al.}, Phys. Rev. Lett. {\bf 122}, 192502 (2019).
\bibitem{Mars2018} B. A. Marsh, T. Day Goodacre, S. Sels, Y. Tsunoda, B. Andel, A. N. Andreyev {\it et al.}, Nature Phys. {\bf 14}, 1163 (2018).
\bibitem{Kuehl1977} T. K\"{u}hl, P. Dabkiewicz, C. Duke, H. Fischer, H.-J. Kluge, H. Kremmling, and E.-W. Otten, Phys. Rev. Lett. {\bf 39}, 180 (1977).
\bibitem{Ulm1986} G. Ulm, S. K. Bhattacherjee, P. Dabkiewicz, G. Huber, H.-J. Kluge, T. Kuehl {\it et al.}, Z. Phys. A {\bf 325}, 247 (1986). 
\bibitem{Anse1986} M. Anselment, W. Faubel, S. Goering, A. Hanser, G. Meisel, H. Rebel and G. Schatz, Nucl. Phys. A {\bf 451}, 471 (1986).
\bibitem{Barz2021} A. Barzakh, A. N. Andreyev, C. Raison, J. G. Cubiss, P. Van Duppen, S. P\'{e}ru {\it et al.} Phys. Rev. Lett. {\bf 127}, 192501 (2021).
\bibitem{Godd2013} P. M. Goddard, P. D. Stevenson, A. Rios, Phys. Rev. Lett. {\bf 110}, 032503 (2013).
\bibitem{Hamm2018} M. Hammen, W. Nörtershäuser, D.L. Balabanski et al.,. Phys. Rev. Lett. {\bf 121}, 102501 (2018).
\bibitem{Repo2021} M. Reponen, R.P. de Groote, L. Al Ayoubi et al., Nat. Commun. {\bf 12}, 4596 (2021).
\bibitem{Malb2022} S. Malbrunot-Ettenauer, S. Kaufmann, S. Bacca et al., Phys. Rev. Lett. {\bf 128}, 022502 (2022). 
\bibitem{Sun2017} B. H. Sun, C. Y. Liu, H. X. Wang, Phys. Rev. C {\bf 95}, 014307 (2017). 
\bibitem{CMa2021} C. Ma, Y. Y. Zong, Y. M. Zhao, and A. Arima, Phys. Rev. C {\bf 104}, 014303 (2021).
\bibitem{Buch2005} F. Buchinger, J. M. Pearson, Phys. Rev. C {\bf 72}, 057305 (2005).
\bibitem{Iimu2008} H. Iimura, F. Buchinger, Phys. Rev. C {\bf 78}, 067301 (2008).
\bibitem{Gori2010} S. Goriely, N. Chamel, J.M. Pearson, Phys. Rev. C {\bf 82}, 035804 (2010). 
\bibitem{Rein2017} P.-G. Reinhard, W. Nazarewicz, Phys. Rev. C {\bf 95}, 064328 (2017).
\bibitem{ADND2022} K. Zhang, M.-K. Cheoun, Y.-B. Choi, P. S. Chong, J. Dong, Z. Dong {\it et al.}, At. Data Nucl. Data Tables {\bf 144}, 101488 (2022).
\bibitem{Xia2018} X. W. Xia, Y. Lim, P. W. Zhao, H. Z. Liang, X. Y. Qu, Y. Chen {\it et al.}, At. Data Nucl. Data Tables {\bf 121}, 1 (2018).
\bibitem{Pere2021} U. C. Perera, A. V. Afanasjev, P. Ring, Phys. Rev. C {\bf 104}, 064313 (2021). 
\bibitem{Fors2009} C. Forss\'{e}n, E. Caurier, P. Navr\'{a}til, Phys. Rev. C {\bf 79}, 021303(R) (2009).
\bibitem{Chou2020} P. Choudhary, P. C. Srivastava, P. Navr\'{a}til, Phys. Rev. C {\bf 102}, 044309 (2020).
\bibitem{YFMa2020} Y. F. Ma, C. Su, J. Liu {\it et al.}, Phys. Rev. C {\bf 101}, 014304 (2020).
\bibitem{Wu2020} D. Wu, C. L. Bai, H. Sagawa {\it et al.}, Phys. Rev. C {\bf 101}, 054323 (2020).
\bibitem{JQMa2022} J. Q. Ma and Z. H. Zhang, Chin. Phys. C {\bf 46}, 074105 (2022).
\bibitem{Dong2023} X.-X. Dong, R. An, J.-X. Lu, and L.-S. Geng, Phys. Lett. B {\bf 838}, 137726 (2023).
\bibitem{Mun2023} Myeong-Hwan Mun, Seonghyun Kim, Myung-Ki Cheoun, W. Y. So, Soonchul Choi, and Eunja Ha, Phys. Lett. B {\bf 847}, 138298 (2023)
\bibitem{Sels2019} S. Sels, T. Day Goodacre, B. A. Marsh, A. Pastore, W. Ryssens, Y. Tsunoda {\it et al}.,  Phys. Rev. C {\bf 99}, 044306 (2019).
\bibitem{Zhou2010} S.-G. Zhou, J. Meng, P. Ring, and E.-G. Zhao, Phys. Rev. C {\bf 82}, 011301(R) (2010).
\bibitem{Lulu2012} L. Li, J. Meng, P. Ring, E.-G. Zhao, and S.-G. Zhou, Phys. Rev. C {\bf 85}, 024312 (2012).
\bibitem{Li2012} L.-L. Li, J. Meng, P. Ring, E.-G. Zhao, and S.-G. Zhou, Chin. Phys. Lett. {\bf 29}, 042101 (2012).
\bibitem{Kai2020} K. Zhang, M.-K. Cheoun, Y.-B. Choi, P. S. Chong, J. Dong, L. Geng {\it et al.}, Phys. Rev. C {\bf 102}, 024314 (2020).
\bibitem{Cong2022} C. Pan, M.-K. Cheoun, Y.-B. Choi, J. Dong, X. Du, X.-H. Fan {\it et al}., Phys. Rev. C {\bf 106}, 014316 (2022). 
\bibitem{Kaiyuan2021} K. Zhang, X. He, J. Meng, C. Pan, C. Shen, C. Wang, and S. Zhang, Phys. Rev. C {\bf 104}, L021301 (2021).
\bibitem{Pan2021} C. Pan, K. Y. Zhang, P. S. Chong, C. Heo, M. C. Ho, J. Lee {\it et al}.,  Phys. Rev. C {\bf 104}, 024331 (2021).
\bibitem{Peng2024} Peng Guo, Xiaojie Cao, Kangmin Chen, Zhihui Chen, Myung-Ki Cheoun, Yong-Beom Choi {\it et al}., arXiv: 2402.02935 (2024).
\bibitem{Kim2022} S. Kim, M.-H. Mun, M.-K. Cheoun, and E. Ha, Phys. Rev. C {\bf 105}, 034340 (2022).
\bibitem{Cong2019} C. Pan, K. Zhang, and S. Zhang, Int. J. Mod. Phys. E {\bf 28}, 1950082 (2019).
\bibitem{In2021} E. J. In, P. Papakonstantinou, Y. Kim, and S.-W. Hong, Int. J. Mod. Phys. E {\bf 30}, 2150009 (2021).
\bibitem{Sun2021} X.-X. Sun, Phys. Rev. C {\bf 103}, 054315 (2021).
\bibitem{Sun2021-2} X.-X. Sun and S.-G. Zhou, Sci. Bull. {\bf 66}, 2072 (2021).
\bibitem{Sun2021-3} X.-X. Sun and S.-G. Zhou, Phys. Rev. C {\bf 104}, 064319 (2021).
\bibitem{Sun2020} X.-X. Sun, J. Zhao, and S.-G. Zhou, Nucl. Phys. A {\bf 1003}, 122011 (2020).
\bibitem{Sun2018} X.-X. Sun, J. Zhao, and S.-G. Zhou, Phys. Lett. B {\bf 785}, 530 (2018).
\bibitem{Yang2021} Z. H. Yang, Y. Kubota, A. Corsi, K. Yoshida, X.-X. Sun, J. G. Li {\it et al.}, Phys. Rev. Lett. {\bf 126}, 082501 (2021).
\bibitem{Zhang2019} K. Y. Zhang, D. Y. Wang, and S. Q. Zhang, Phys. Rev. C {\bf 100}, 034312 (2019)
\bibitem{Zhang2023} K. Y. Zhang, P. Papakonstantinou, M.-H. Mun, Y. Kim, H. Yan, and X.-X. Sun, Phys. Rev. C {\bf 107}, L041303 (2023)
\bibitem{ZhangPLB} K. Y. Zhang, S. Q. Yang, J. L. An, S. S. Zhang, P. Papakonstantinou, M.-H. Mun {\it et al.},  Phys. Lett. B {\bf 844}, 138112 (2023)
\bibitem{An2024} Jia-Lin An, Kai-Yuan Zhang, Qi Lu, Shi-Yi Zhong, and Shi-Sheng Zhang, Phys. Lett. B {\bf 849}, 138422 (2024).
\bibitem{Peng2023} P. Guo, C. Pan, Y. C. Zhao, X. K. Du, and S. Q. Zhang, Phys. Rev. C {\bf 108}, 014319 (2023).
\bibitem{Kucharek1991} H. Kucharek and P. Ring, Z. Phys. A {\bf 339}, 23 (1991).
\bibitem{Zhou2003} Shan-Gui Zhou, Jie Meng, and P. Ring, Phys. Rev. C {\bf 68}, 034323 (2003).
\bibitem{Dirac2022} K. Y. Zhang, C. Pan, and S. Q. Zhang, Phys. Rev. C {\bf 106}, 024302 (2022).
\bibitem{Ring2004} P. Ring and P. Schuck, \textit{The Nuclear Many-Body Problem} (Springer Science $\And$ Business Media, 2004).
\bibitem{Price1987} C.~Price and G.~Walker, Phys. Rev. C {\bf 36}, 354 (1987).
\bibitem{Martin2008} S. Perez-Martin and L. M. Robledo, Phys. Rev. C {\bf 78}, 014304 (2008).
\bibitem{AME2020} M. Wang, W. J. Huang, F. G. Kondev, G. Audi, and S. Naimi, Chin. Phys. C {\bf 45}, 030003 (2021). 
\bibitem{NNDC} NNDC (National Nuclear Data Center), Brookhaven National Laboratory, https://www.nndc.bnl.gov/nudat2/
\bibitem{Pove2016} J. L. Wood and K. Heyde, J. Phys. G: Nucl. Part. Phys. {\bf 43}, 020402 (2016).
\bibitem{Gade2016} A. Gade and S. N. Liddick, J. Phys. G: Nucl. Part. Phys. {\bf 43}, 024001 (2016).
\bibitem{Heyde2011} K. Heyde and J. L. Wood, Rev. Mod. Phys. {\bf 83}, 1467 (2011), [Erratum Rev. Mod. Phys. {\bf 83}, 1655 (2011)].
\bibitem{Nach2004} E. N\'{a}cher, A. Algora, B. Rubio, J. L. Ta\'{i}n, D. Cano-Ott, S. Courtin {\it et al.}, Phys. Rev. Lett {\bf 92}, 232501 (2004).
\bibitem{Andr2000} A. N. Andreyev, M. Huyse, P. Van Duppen, L. Weissman, D. Ackermann, J. Gerl {\it et al.}, Nature {\bf 405}, 430 (2000).
\bibitem{Ojala2022} Joonas Ojala, Janne Pakarinen, Philippos Papadakis, Juha Sorri, Mikael Sandzelius {\it et al.}, Communications Physics {\bf 5}, 213 (2022). 
\bibitem{Paul2009} P. F. Mantica, Physics {\bf 2}, 18 (2009).
\bibitem{Sava1990} G. Savard, J. Crawford, J. Lee, G. Thekkadath, H. Duong, J. Pinard {\it et al.}, Nucl. Phys. A {\bf 512}, 241 (1990).
\bibitem{Pass1994} G. Passler, J. Rikovska, E. Arnold, H.-J. Kluge, L. Monz, R. Neugart {\it et al.}, Nucl. Phys. A A {\bf 580}, 173 (1994).
\bibitem{Blan1997} F. Le Blanc, J. Obert, J. Oms, J. C. Putaux, B. Roussi\'{e}re, J. Sauvage {\it et al.}, Phys. Rev. Lett. {\bf 79}, 2213 (1997).
\bibitem{Prit2016} B. Pritychenko, M. Birch, B. Singh, M. Horoi, At. Data Nucl. Data Tables {\bf 107}, 1 (2016).
\bibitem{Seli2009} M. D. Seliverstov , A. N. Andreyev, N. Barr\'{e}, A. E. Barzakh, S. Dean, H. De Witte {\it et al.}, Eur. Phys. J. A {\bf 41}, 315 (2009).
\bibitem{Hori2022} W. Horiuchi and T. Inakura, Phys. Rev. C {\bf 105} 044303 (2022); Phys. Rev. C {\bf 101}, 061301(R) (2020).
\bibitem{YLYang2021} Y. L. Yang, Y. K. Wang, P. W. Zhao, and Z. P. Li, Phys. Rev. C {\bf104}, 054312 (2021).
\end{thebibliography}
\end{document}